\documentclass[11pt]{article}

\usepackage[utf8]{inputenc}
\usepackage[T1]{fontenc}
\usepackage[margin=1in]{geometry}
\usepackage{amsmath}
\usepackage{amssymb}
\usepackage{graphicx}
\usepackage{subcaption}
\usepackage{booktabs}
\usepackage{placeins}

\usepackage[blocks]{authblk}

\usepackage{hyperref}

\graphicspath{{image/}}

\title{FaceKit: a Toolkit for Interpretable Facial Phenotyping, Synthetic Image Generation and Privacy Analysis in Rare Diseases}

\author[1,2]{Hongzhuo Chen}
\author[1,2]{Zhanliang Wang}
\author[3,4]{Florent Pollet}
\author[1]{Mian Umair Ahsan}
\author[5]{Joshua Bie}
\author[6]{Tzung-Chien Hsieh}
\author[6]{Peter Krawitz}
\author[7]{Cong Liu}
\author[7]{Wendy K Chung}
\author[3]{Chunhua Weng}
\author[3,4,8]{Gamze Gürsoy}
\author[1,9,*]{Kai Wang}

\affil[1]{Raymond G. Perelman Center for Cellular and Molecular Therapeutics,
Children's Hospital of Philadelphia, Philadelphia, PA, USA}

\affil[2]{Applied Mathematics and Computational Science Graduate Program,
University of Pennsylvania, Philadelphia, PA, USA}

\affil[3]{Department of Biomedical Informatics, Columbia University,
New York, NY, USA}

\affil[4]{New York Genome Center, New York, NY, USA}

\affil[5]{Havard Westlake School, Studio City, CA, USA}

\affil[6]{Institute for Genomic Statistics and Bioinformatics, University Hospital Bonn, Rheinische Friedrich-Wilhelms-Universitat Bonn, Bonn, Germany}

\affil[7]{Department of Pediatrics, Boston Children's Hospital \& Harvard Medical School, Boston, MA, USA}

\affil[8]{Department of Computer Science, Columbia University,
New York, NY, USA}

\affil[9]{Department of Pathology and Laboratory Medicine,
Perelman School of Medicine, University of Pennsylvania,
Philadelphia, PA, USA}

\affil[*]{Corresponding author:
\href{mailto:wangk@chop.edu}{\texttt{wangk@chop.edu}}}

\date{}

\begin{document}

\maketitle

\begin{abstract}
 Many rare genetic diseases are associated with recognizable craniofacial features. However, traditional approaches for describing facial morphology rely largely on qualitative clinical observation and free-text descriptions, which are often subjective, non-standardized, and difficult to reproduce across observers and institutions. Although the Human Phenotype Ontology (HPO) provides controlled terms for describing facial features, these terms are typically categorical rather than quantitative and may vary depending on examiner experience and interpretation. Here, we present FaceKit, a computational framework for quantitative facial phenotyping from frontal facial photographs. FaceKit extracts standardized measurements of facial landmarks and derived 120 morphological features, then reports feature-level z-scores representing deviation from population reference distributions. The reference distributions are built from the FairFace dataset spanning diverse ancestral groups. We evaluated FaceKit on a curated subset of the GestaltMatcher Database covering 50 rare-disease cohorts. In addition to quantitative facial analysis, FaceKit includes synthetic facial image generation to support rare disease model development and data augmentation. We also performed privacy evaluation to assess whether synthetic images reveal identifiable information from real patient photographs and could compromise patient privacy. Across disease case studies, FaceKit-derived quantitative measurements captured known facial features associated with rare genetic disorders and provided objective support for clinical phenotyping.
Together, these results establish FaceKit as a useful tool for quantitative phenotyping, and has the potential to improve rare disease diagnosis, support genotype-phenotype studies, and enable more reproducible clinical characterization across diverse patient populations.
\end{abstract}


\section{Introduction}
Facial morphology is an important component of phenotypic characterization in many rare genetic diseases and can provide useful phenotypes for disease recognition alongside other clinical and genetic information. Traditionally, these patterns have been assessed through expert observation and described using standardized phenotype terminology such as the Human Phenotype Ontology (HPO). More recently, neural network \cite{DeepGestalt,hsieh2022gestaltmatcher} and vision language models \cite{mint, gestaltmml} have also used facial images to support rare-disease recognition and the identification of facial phenotype terms.

However, existing approaches do not fully address the need for quantitative and interpretable facial phenotyping. HPO terms provide a standardized vocabulary, but many facial features are still evaluated qualitatively and are not directly linked to reproducible measurements from images. Similarly, learning-based models commonly produce disease rankings, probabilities, or phenotype labels, while the facial characteristics underlying these outputs are often difficult to express.  General-purpose facial landmark tools provide image coordinates, but researchers must still define their own normalization procedures, geometric measurements, feature-selection rules, and output formats. This fragmentation makes it difficult to compare facial morphology consistently across patients, diseases, and study cohorts.

To address this gap, we developed FaceKit, a lightweight and interpretable toolkit that converts facial images into structured quantitative representations of facial morphology. FaceKit integrates facial landmark extraction and visualization, frontal-pose quality filtering, geometric normalization, and cohort average-face generation. It computes 120 interpretable measurements across major facial regions, reports each of them as a $z$-score against a normative reference distribution distributed with the toolkit, and supports disease-aware feature selection for downstream cohort analysis. FaceKit also provides an extensible framework through which researchers can define additional geometric features while maintaining a unified output representation. Together, these functions provide a reproducible pipeline for transforming facial images into quantitative phenotypic features for rare-disease research.

\section{Methods}

\begin{figure}[!htbp]
  \centering
  \includegraphics[width=\linewidth]{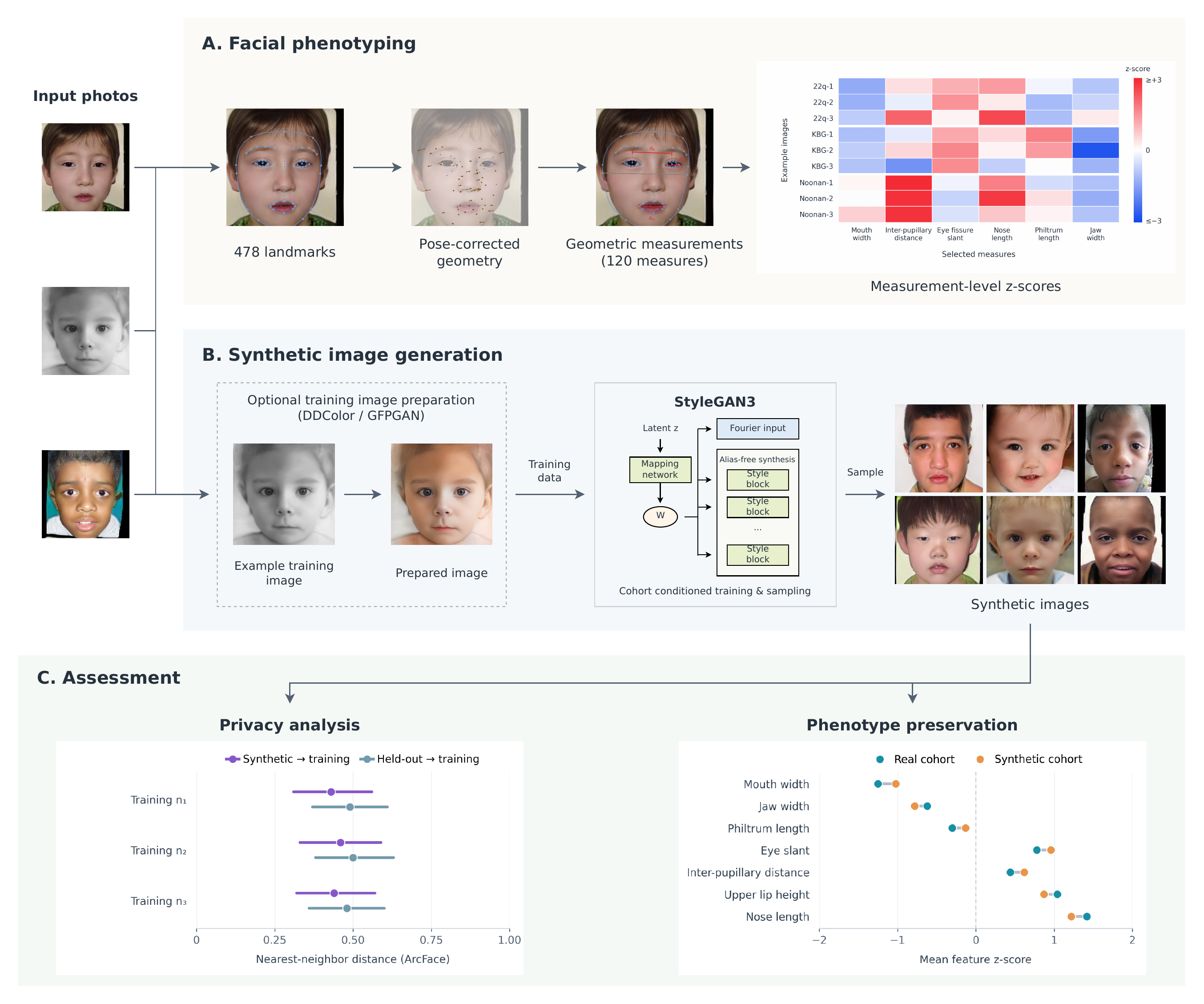}
  \caption{\textbf{Overview of the FaceKit workflow.}
  (A) Facial phenotyping: 478 landmarks are extracted from each photograph and
  corrected for head pose, 120 geometric measurements are derived, and each is
  expressed as a $z$-score relative to a normative reference population.
  (B) Synthetic image generation: after optional preparation with DDColor and
  GFPGAN, the training images are used to fit a StyleGAN3 generator conditioned
  on the cohort label, from which synthetic faces of each cohort are sampled.
  (C) Assessment: privacy is evaluated by comparing synthetic-to-training with
  held-out-to-training nearest-neighbor distances across cohorts of different
  training size, and phenotype preservation by comparing measurements between
  real and synthetic cohorts. All facial images shown are synthetic, and the
  assessment plots are schematic.}
  \label{fig:facekit-workflow}
\end{figure}

FaceKit provides an integrated workflow for rare disease facial image datasets that spans four stages: image preprocessing, landmark-based quantitative phenotyping, synthetic image generation, and privacy assessment of the resulting generative models (Fig.~\ref{fig:facekit-workflow}). The workflow begins with image preprocessing and landmark-based facial phenotyping. This stage generates standardized facial images, landmark representations, and interpretable geometric measurements which can be used for downstream machine learning tasks and generative model development. Building on these standardized images, FaceKit supports the training of cohort-conditioned generative models to produce synthetic facial images for data augmentation and methodological evaluation. The generated images can subsequently be processed through the same phenotyping workflow to evaluate whether they preserve the corresponding geometric characteristics. Finally, because generative models trained on patient facial images may inadvertently retain or reproduce information specific to the training data, FaceKit also includes a privacy-assessment framework to evaluate whether generated images reproduce identifiable information or closely resemble the real patient photographs used for training. The study is approved by the institutional review board of the Children's Hospital of Philadelphia.

\subsection{Image Preprocessing}

FaceKit accepts facial images provided either in a single directory or in a hierarchical directory structure organized by disease cohort. Each image is first processed using MediaPipe Face Landmarker to detect 478 three-dimensional facial landmarks and estimate a 4 $\times$ 4 facial transformation matrix describing the rigid pose of the head relative to the camera. The transformation matrix is converted into yaw, pitch, and roll angles to assess whether the image is sufficiently frontal for subsequent geometric analysis. By default, images with an absolute yaw or pitch greater than $15^{\circ}$, or an absolute roll greater than $10^{\circ}$, are marked as non-frontal. This quality-control procedure reduces variation introduced by head pose and identifies images for which facial measurements may be unreliable. For each image, FaceKit stores the landmark coordinates together with the image identifier and dimensions. The extracted data can be saved as individual JSON files or combined into a JSONL file for batch processing and downstream phenotype extraction. FaceKit additionally provides visualization utilities that overlay the detected landmarks on the original image, either as a connected mesh or as individual points with selected landmark groups, allowing landmark placement to be inspected manually where detection quality is in question.



\subsection{Geometric Phenotype Extraction}

FaceKit converts the extracted facial landmarks into structured quantitative
representations of facial morphology. Before feature calculation, landmark
coordinates are geometrically normalized to reduce variation caused by image
position, orientation, and scale.

Head pose is removed first. Under an orthographic camera, a landmark lying at
$X$ along the face plane and at depth $Z$ appears at horizontal image position
$x_{\mathrm{img}} = X\cos\theta + Z\sin\theta$ when the head is turned by
$\theta$ (Fig.~\ref{fig:posecorr-arms}\subref{fig:posecorr-geom}). The first
term shrinks the face horizontally by a factor shared by every landmark; the
second displaces each landmark in proportion to how deep it sits, and it is the
only term that requires a depth. FaceKit applies the same correction for an
arbitrary head rotation $R$, with the three pose angles acting together rather
than yaw alone, recovering the frontal in-plane coordinates as

\begin{equation}
\begin{pmatrix} X \\ Y \end{pmatrix}
  = \bigl(R_{[:2,\,:2]}\bigr)^{-1}
    \left[ \begin{pmatrix} x \\ y \end{pmatrix} - R_{[:2,\,2]}\, Z \right],
\label{eq:derot}
\end{equation}


where $R_{[:2,:2]}$ is the part of the rotation that maps the in-plane position
into the image and $R_{[:2,2]}$ the part that carries the depth, subtracted
before the inversion. $R$ is the upper-left $3\times3$ rotation block of the
$4\times4$ MediaPipe facial transformation matrix, replaced by its nearest pure
rotation through singular-value decomposition, since the fitted block carries
residual scale and shear that the inversion would otherwise amplify.

A depth $Z$ required by equation~\eqref{eq:derot} can be directly obtained from MediaPipe. Each landmark carries a $z$ coordinate measured
from a plane through the center of mass of the mesh and scaled to the same units
as the others. However, the z coordinate is not obtained in the same
way as the first two. During training, the $x$ and $y$ coordinates were
iteratively refined against human annotation of real photographs, while the $z$
coordinates were left untouched and supervised only by synthetic renderings of a
morphable face model. In addtion, the depth is reported by its authors as not metrically accurate, and the model is evaluated on two-dimensional
error alone~\cite{mediapipe_facemesh}. Because a morphable model can only represent faces within the linear span of the population it was built from, the depth is expected to stay close to the mean profile of that population, and the output behaves accordingly: across the 4,559 images analyzed here, the per-image depth profile correlates with the mean profile over all images at a median $r = 0.94$, and only 17.8\% of its variance is specific to the image from which it was computed.


FaceKit therefore applies the same depth to every image. The value assigned to a
landmark comes from a canonical-face table of 478 constants, computed as the mean
de-rotated depth over the images analyzed here.
Averaging also discards the per-image fluctuation of the detector's $z$, which,
given how that coordinate was supervised, is where the fitting error of any
single photograph resides. We also tested on two alternatives: applying no depth at
all, by setting $Z = 0$ and treating the face as flat, or applying each image's
own $z$ as returned by the detector. Both are compared against the canonical
table and against uncorrected landmarks in Section~\ref{sec:posecorr}, on
synthetic faces where the true depth is known and on repeat photographs of the
same patients, and the canonical table is the one that performs best.

The pose-corrected configuration is then brought to a common frame. It is
translated so that the midpoint of the inner canthi lies at the origin, and every
measurement is expressed as a ratio to an anatomically defined reference length
rather than in pixels: bizygomatic width, facial height, midface height or mouth
width, chosen according to the facial region and the quantity being measured.
Expressing the measurements as ratios is what makes them comparable across
photographs taken at different distances and resolutions, and it also determines
how much of the pose problem survives the correction.

Using the normalized landmarks, FaceKit calculates 120 interpretable geometric features describing regional and global facial morphology. These features include distances, ratios, angles, thickness measurements, curvature--related measurements, asymmetry indices, and facial proportions across the eyebrows, eyes, nose, philtrum, mouth and lips, chin and jaw, forehead, midface and cheeks, and the overall face. Each feature is defined using explicit landmark combinations and geometric operations, allowing its value to be directly inspected and compared across individuals or cohorts.


\subsection{Normative scoring}

A geometric measurement expressed in its own units does not by itself indicate
whether a face is unusual. FaceKit therefore reports every measurement as a
$z$-score against a normative reference,

\begin{equation}
z = \frac{x - \mu_{\mathrm{ref}}}{\sigma_{\mathrm{ref}}},
\end{equation}

where $\mu_{\mathrm{ref}}$ and $\sigma_{\mathrm{ref}}$ are the mean and standard
deviation of that measurement in a control population. A value is then read as
the number of reference standard deviations by which a face departs from
controls, which places measurements with different definitions and different
natural scales on a common axis and makes them comparable to one another.

The reference distributed with FaceKit was built from 2,051 FairFace images
sampled across three ancestry groups, of which 886 pass the same frontal-pose
gate that is applied to the faces being scored. Applying the gate on both sides
is deliberate: an asymmetry or slant measurement responds to whatever residual
head pose survives the gate, so a reference measured under looser acquisition
conditions than the faces scored against it contributes an inflated standard
deviation and understates their deviation. Users may supply their own control
images instead, from which a reference is derived in the same way. Because the
reference is pooled over ancestry groups rather than conditioned on them, a
$z$-score expresses departure from the control population as a whole. A
reference and the faces scored against it must be extracted under the same
convention, since the pose correction described above fixes the meaning of all
120 measurements simultaneously.

\subsection{Average-face generation}

FaceKit generates average-face representations to summarize shared facial morphology within an image cohort. The implementation incorporates source code from the FaceMorpher software~\cite{facemorpher}. Facial images are first aligned to a common landmark configuration to reduce variation in position, orientation, and scale, after which the aligned images are combined into a cohort-level average. Users can specify the output resolution and optionally apply edge smoothing, a transparent background, or blending with a destination face. The resulting average face provides an intuitive visual summary of common facial patterns within a cohort and can facilitate qualitative comparisons across disease groups.

\subsection{Synthetic Facial Image Generation}

Generative models have previously been used to synthesize condition-specific facial portraits for rare genetic disorders~\cite{gestaltgan}. FaceKit additionally supports the generation of synthetic facial images.
Two usage modes are provided. Users with a sufficiently large image
collection can train a generator on their own cohorts and sample from the
resulting weights, while other users can sample directly from a generator
pre-trained on the 50 rare-disease cohorts of this study and distributed with
FaceKit.

Facial photographs collected for rare-disease research are often
retrospective and heterogeneous in acquisition quality, and the images used
for training are therefore standardized beforehand. Images that cannot be
used for morphological modelling may optionally be excluded at this stage,
including those obscured by privacy blocks, those of extremely low quality,
and those in which facial features are not clearly visible. Grayscale or
colour-degraded photographs are then colorized using DDColor~\cite{ddcolor},
and blurred or low-resolution faces are restored and upsampled using
GFPGAN~\cite{gfpgan}. Finally, FaceKit locates the facial region from the
detected landmarks, removes the surrounding background, and resizes the
retained facial area to $256\times256$ pixels, so that all training images
share a common facial framing and resolution. This preparation applies to
the training corpus only and is not required when sampling from an existing
generator. The GMDB images used for the pre-trained generator were already
restored face crops, and were therefore only resized from $224\times224$ to
$256\times256$ pixels, without the landmark-based re-cropping.

\begin{figure}[!htbp]
  \centering
  \includegraphics[width=\linewidth]{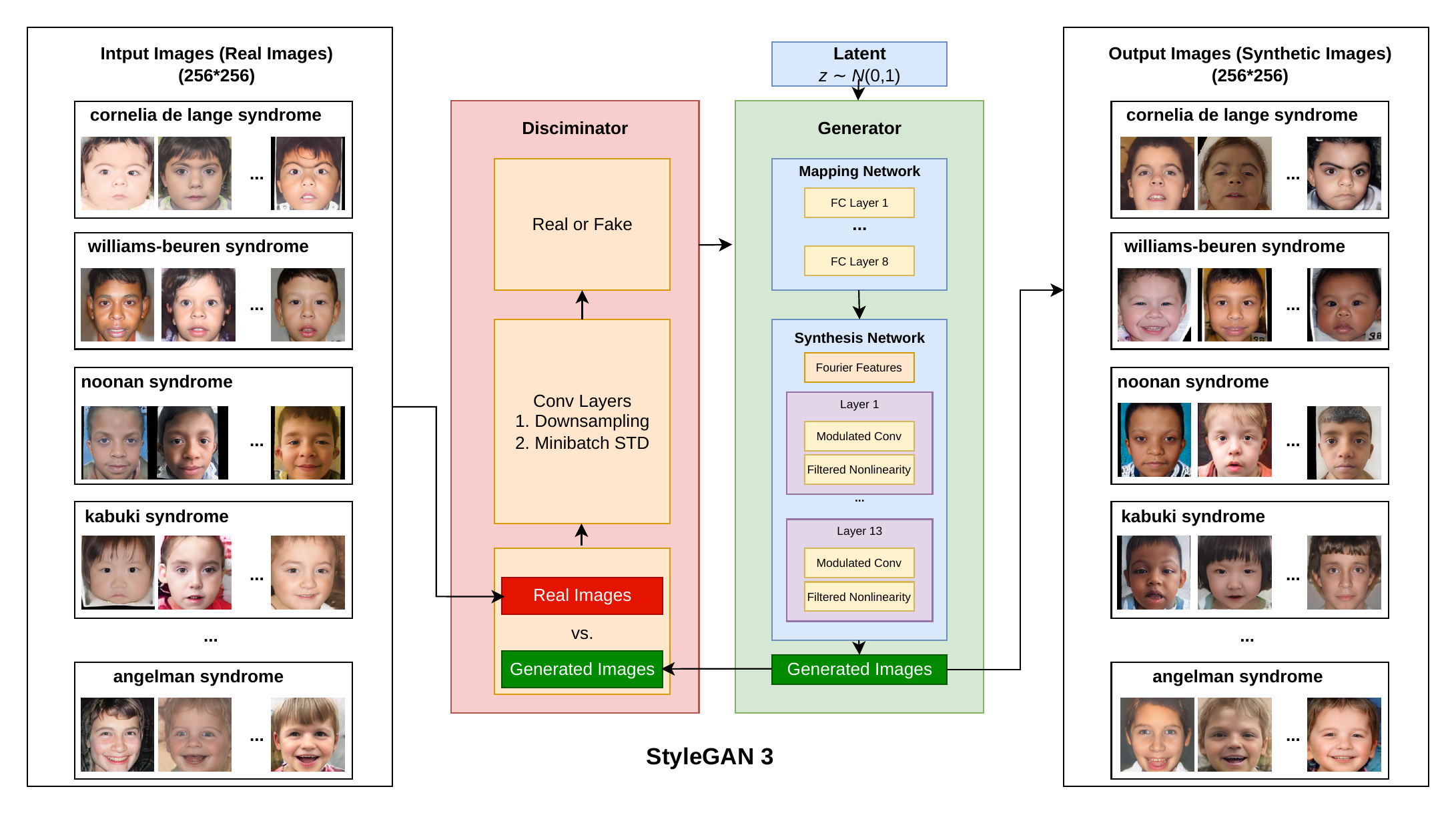}
  \caption{\textbf{Synthetic facial image generation.} Prepared real images of
  the rare-disease cohorts (left) are used, together with their cohort labels,
  to train a conditional StyleGAN3 generator. The
  discriminator applies successive downsampling convolutions with a minibatch
  standard-deviation layer and separates real from generated images, while the
  generator maps a latent vector $z \sim \mathcal{N}(0, I)$ and the embedded
  cohort label through an eight-layer mapping network into a style vector that
  modulates the convolutions of the synthesis network. Sampling from the
  trained generator with a given label yields $256\times256$ synthetic facial
  images of that cohort (right). All faces shown in this schematic are
  synthetic examples; those in the left panel represent the real training
  inputs.}
  \label{fig:syngen}
\end{figure}

Generators are trained using StyleGAN3~\cite{stylegan3}. A latent vector
$z \sim \mathcal{N}(0, I)$ is transformed by a mapping network into an
intermediate style vector, which modulates the convolutions of a synthesis
network to produce an image, while a discriminator is trained in parallel to
distinguish real from generated images; the two networks are optimized
adversarially (Fig.~\ref{fig:syngen}). A single generator is trained for all
cohorts of a collection and is conditioned on the cohort label, which is
embedded and passed to the mapping network together with $z$ and is also
supplied to the discriminator, so that sampling can be directed to any cohort.
The pre-trained generator was fitted to all 4,559 prepared images of the 50
cohorts, between 45 and 372 images per cohort, with the StyleGAN3-T
configuration at $256\times256$ resolution, an R1 regularization weight of
$\gamma = 2$, horizontal-flip augmentation in addition to adaptive
discriminator augmentation, a batch size of 32 and a training length of
5{,}000 kimg. The generator and discriminator learning rates are initialized
at $2.5\times10^{-3}$ and $2\times10^{-3}$, respectively, and follow a
cosine decay schedule. Synthetic images of a cohort are then obtained by
drawing latent vectors from the standard normal distribution, without
truncation, and passing them together with the cohort label through
either a user-trained or a pre-trained generator.

\subsection{Phenotype preservation of synthetic images}

Whether a generator preserves the geometric phenotype of its training cohorts is
assessed by passing real and synthetic images through the identical FaceKit
pipeline. For each cohort, as many synthetic images are sampled as the cohort
has real training images, and each cohort is restricted to the measurements to which its
HPO annotations map(Section~\ref{sec:validity}); a cohort without a mapped
measurement is never analyzed as the focal cohort but remains part of the
comparison pool. Within each cohort, the synthetic and real distributions of a
measurement are compared by Cohen's $d$ with a pooled standard deviation,
provided both sets contain at least five values. To test whether the synthetic
images preserve what distinguishes one cohort from the others, the effect of
the focal cohort against all remaining cohorts pooled is computed as Cohen's
$d$, separately in the synthetic and in the real images. The two effects are
compared by whether their signs agree and, for every cohort with at least three
measurements, by the Spearman and Pearson correlations between the two effect
vectors. Dispersion is compared as the ratio of the synthetic to the real
variance of each measurement within a cohort, summarized by facial region.

\subsection{Privacy assessment of the pre-trained generator}

Because the generator distributed with FaceKit is trained on photographs of
real individuals, the synthetic images it produces are assessed before release
for residual similarity to the training data. Two aspects are examined
separately: whether a synthetic image reproduces the identity of a training
individual, and whether it reproduces the appearance of a training photograph.

For the identity assessment, the prepared images of each cohort are divided into
a training partition and a held-out partition that is disjoint from it at the
patient level. We used the patient-level 80/20 split of the 50 cohorts applied
throughout this study (random seed 42), which gives 3,602 training images of
2,547 patients and 957 held-out images of 661 patients. Because the released
generator is trained on both partitions, the audit is carried out on a second
generator with the same architecture and training settings that is fitted to
the training partition alone, so that the held-out images never enter its
training. From this audited generator, five synthetic images are sampled for
every held-out image of each cohort, conditioned on that cohort's label.
Faces are detected with InsightFace~\cite{insightface}, and the five
detected keypoints are used to align each face to a $112\times112$ crop through
a similarity transform. Identity embeddings are extracted with
ArcFace~\cite{arcface}, and the entire procedure is repeated with
AdaFace~\cite{adaface}, using an IR-50 backbone trained on MS1MV2, and with
LVFace~\cite{lvface}, using the LVFace-L model trained on Glint360K, so that the
assessment does not depend on a single recognition model; images in which no
face is detected are excluded. Distances between embeddings are measured as
cosine distances, and every synthetic and every held-out image is characterized
by its distance to the nearest training image of the same cohort. Rather than
fixing a verification threshold in advance, the threshold is read off the
held-out distribution: for a given percentile $p$, it is set to the $p$-th
percentile of the held-out nearest-neighbor distances, so that the operating
point corresponds to a fixed false-acceptance rate among real images. The
proportion of synthetic images falling below this threshold is recorded at
$p = 1$, $2$, $5$, $10$ and $20$. Every nearest-neighbor search is confined to
a single cohort: a synthetic or a held-out image is compared only with the
training images of the cohort it was drawn from, and no distance is computed
between cohorts, so that a resemblance between two patients who share a syndrome
is never mistaken for one between cohorts. What is pooled across cohorts is the
resulting set of per-image distances, from which one threshold per percentile is
read. The ordinal behaviour of each
embedding model is verified on the same images: same-person pairs are formed
from images of one patient acquired no more than half a year apart, so that the
within-patient distance reflects what the recognition embedding does with one
identity rather than the shape change a growing child undergoes between two
sessions, and each such pair is compared against 30 randomly drawn images of
other patients from the same cohort and partition, checking whether the
same-person distance is the smaller one throughout. The different-person
distance distribution is formed from every pair of images of two different
patients within one cohort and partition.

The appearance assessment follows the same design with a perceptual rather than
an identity metric. LPIPS distances~\cite{lpips} are computed with an AlexNet
backbone on images resized to $256\times256$, and each synthetic and held-out
image is compared exhaustively against every training image of its cohort,
retaining the smallest distance; thresholds are again taken from the percentiles
of the held-out distribution. The same-person and different-person distributions
are formed again with this metric, but over every available pair of images of one
patient rather than over the half-year window used above: the window is there to
keep facial growth out of an identity comparison, and LPIPS is not read as an
identity metric. Different-person pairs take one randomly chosen image of each
of two patients, with at most 200 pairs per cohort and partition. The two metrics therefore rest on different numbers of
same-person pairs, which is stated where the counts appear. Both assessments are finally summarized by the
nearest neighbor adversarial accuracy~\cite{nnaa}. For two sets of images $A$
and $B$ under a distance $d$, write $d_{AB}(a) = \min_{b \in B} d(a, b)$ for the
distance from $a$ to its nearest neighbor in the other set and
$d_{AA}(a) = \min_{a' \in A \setminus \{a\}} d(a, a')$ for the distance to its
nearest neighbor within its own set. The adversarial accuracy is then
\[
  \mathrm{AA}(A, B) = \frac{1}{2}\left(
    \frac{1}{|A|} \sum_{a \in A} \mathbf{1}\!\left[ d_{AB}(a) > d_{AA}(a) \right]
    + \frac{1}{|B|} \sum_{b \in B} \mathbf{1}\!\left[ d_{BA}(b) > d_{BB}(b) \right]
  \right),
\]
the proportion of images whose nearest neighbor in the other set is farther away
than their nearest neighbor within their own set, averaged over the two
directions; a value of one half indicates that the two sets cannot be
distinguished by proximity, and a value below one half that they lie closer to
each other than to themselves. Writing $S$ for the synthetic images, $T$ for the
training partition and $H$ for the held-out partition, the privacy loss is the
difference
\[
  \Delta = \mathrm{AA}(H, S) - \mathrm{AA}(T, S),
\]
which is positive when the synthetic images lie closer to the training data than
held-out real images of the same cohorts do. Unlike the flagging analysis above,
$\Delta$ is evaluated on the images of all cohorts taken together, so a nearest
neighbor may fall in another cohort; it is computed with embedding distances on
random subsets of 1,000 training and 1,000 synthetic images together with all
held-out images, and with LPIPS distances on 200 images per set, and 95\%
confidence intervals are obtained from 1,000 bootstrap replicates that resample
images with replacement.

Finally, to examine whether cohorts with few training patients are more exposed,
the identity comparison is repeated cohort by cohort without a threshold. For
each cohort and recognition model, the statistic is the probability that a
synthetic image lies closer to the cohort's training partition than a held-out
image of the same cohort does, counting ties as one half, so that one half
indicates no excess. Its association with the number of training patients is
tested across cohorts by the Spearman correlation, with a two-sided permutation
$p$-value from 10,000 permutations. Cohorts are also grouped into six tiers by
the number of training patients (100 or more, 60 to 99, 40 to 59, 30 to 39, 20 to
29, and 13 to 19); within a tier the statistic is pooled over within-cohort
comparisons only, and its 95\% confidence interval is obtained from 1,000
bootstrap replicates that resample held-out images by patient and synthetic
images by image within each cohort.

\section{Results}

We evaluated FaceKit along the major components of its workflow. First, we assessed whether the landmark-derived geometric measurements captured facial phenotypes represented by clinically annotated HPO terms and whether these measurements supported interpretable comparisons across patients and disease cohorts. Second, because every measurement is defined on pose-corrected landmarks, we compared the available ways of supplying the depth that the correction requires, using endpoints that need no annotation. Third, because a measurement is only useful if it survives a change of source, we repeated the disease-separation question on an independent rare-disease image collection and quantified how far the catalogue can be pushed down in image resolution before the measurements shift systematically. Fourth, since every measurement is reported relative to a distributed normative reference, we asked whether conditioning that reference on age, on image resolution or on ancestry changes the conclusions, and whether the deviation a syndrome produces is itself ancestry-dependent, using an independent clinical cohort of 22q11.2 deletion syndrome in which cases and controls are available in every ancestry group. Fifth, we examined whether the cohort-conditioned generator preserved the geometric phenotype distributions of the real images used for training, both within individual disease cohorts and in the differences that distinguish one cohort from another. Finally, because the pre-trained generator is intended for distribution and reuse, we evaluated whether its synthetic outputs reproduced the identities or visual appearance of the real patient photographs used during training. Together, these analyses assess the phenotypic validity, measurement reliability, generative fidelity, and privacy characteristics of the FaceKit workflow.

\subsection{Quantitative measurements reproduce annotated facial phenotypes}
\label{sec:validity}

We applied FaceKit to a curated subset of the GestaltMatcher Database comprising
4,559 frontal facial images from 50 rare-disease cohorts. Although the quality and completeness of Human Phenotype Ontology (HPO) annotations vary across the database, and the annotations may not capture all facial features visible in individual images, this dataset provides a valuable resource for evaluating FaceKit. For each image, FaceKit
extracted 478 facial landmarks and computed 120 unitless geometric measurements.
After pose filtering and collapsing each patient to the median of their frontal
images, 2,897 patients remained. To express every measurement on a common,
clinically interpretable scale, we built a normative reference from 2,051 FairFace
control images spanning three ancestry groups, of which 886 passed the same pose
filter, and converted each measurement into a $z$-score relative to that reference.
Each of the 97 facial HPO terms in the FaceKit vocabulary is mapped to one
measurement together with the direction in which that measurement is expected to
change, so a signed $z$-score can be read directly as agreement or disagreement
with the annotation.

Of the 97 terms, 84 were carried by at least one patient with a gold-standard
annotation; the remaining 13 had no annotated patient anywhere in the cohort and
could not be assessed. Across the 84 assessable terms, the mapped measurement
deviated from the healthy reference in the direction predicted by HPO for 62
(73.8\%), with a mean absolute deviation of 0.47
reference standard deviations. Direction alone does not establish that a deviation is real, so the 78 terms
carried by enough patients to admit a statistical test were examined separately.
Of these, 39 separated from the reference after Benjamini--Hochberg correction,
33 in the direction HPO predicts and 6 against it, while the remaining 39 did
not separate from the reference at all. Agreement was not driven by the terms
carrying the most annotation: terms carried by 10--29 patients agreed as often
as those carried by 30 or more (82.4\% versus 79.5\%).

\begin{figure}[!htbp]
  \centering
  \begin{subfigure}[b]{0.27\linewidth}
    \centering
    \includegraphics[width=\linewidth]{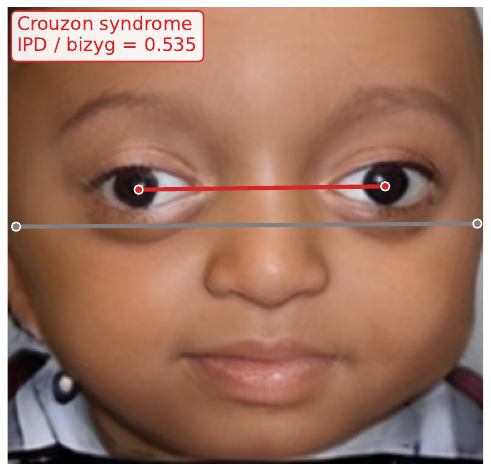}
    \caption{}
    \label{fig:ipd-crouzon}
  \end{subfigure}
  \hfill
  \begin{subfigure}[b]{0.27\linewidth}
    \centering
    \includegraphics[width=\linewidth]{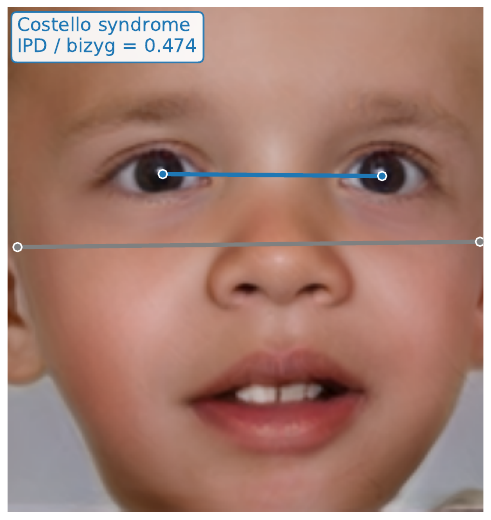}
    \caption{}
    \label{fig:ipd-costello}
  \end{subfigure}
  \hfill
  \begin{subfigure}[b]{0.4\linewidth}
    \centering
    \includegraphics[width=\linewidth]{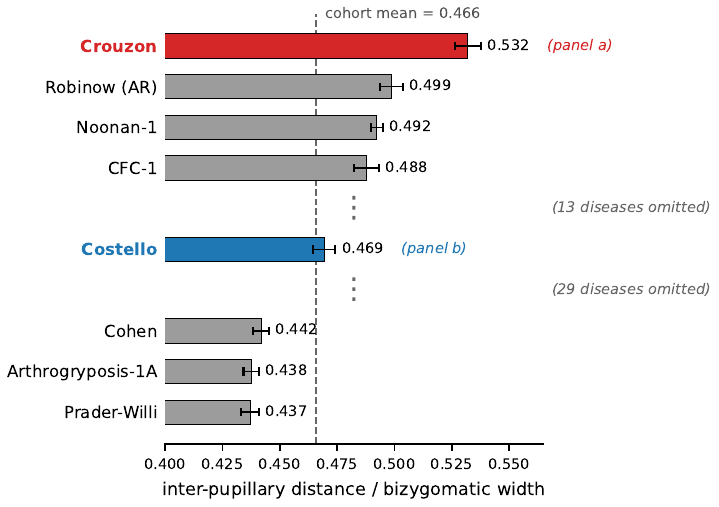}
    \caption{}
    \label{fig:ipd-bars}
  \end{subfigure}

  \caption{\textbf{A single FaceKit measurement, from image to cohort.}
  (\subref{fig:ipd-crouzon}, \subref{fig:ipd-costello}) The inter-pupillary
  distance normalized by bizygomatic width, drawn on the images it is computed
  from: the coloured line joins the two iris centers (landmarks 468 and 473) and
  the grey line joins the two zygomatic points (landmarks 234 and 454); the
  measurement is the ratio of their lengths. (\subref{fig:ipd-crouzon}) A synthetic
  face generated for Crouzon syndrome, the cohort with the highest mean value.
  (\subref{fig:ipd-costello}) A synthetic face generated for Costello syndrome, a
  cohort at the center of the same distribution, shown for contrast. Both faces are
  sampled from the cohort-conditioned generator and depict no real patient.
  (\subref{fig:ipd-bars}) Mean normalized inter-pupillary distance per disease
  cohort. Bars give the cohort mean with its standard error; the dashed line marks
  the mean over the remaining cohorts. Of the 50 cohorts, the four highest, the
  three lowest and Costello syndrome are shown, and the rest are omitted as
  indicated. Values are computed on images passing the frontal-pose filter.}
  \label{fig:ipd}
\end{figure}

To illustrate how the quantitative facial phenotyping module operates, we used hypertelorism in Crouzon syndrome as a case study. Hypertelorism is represented in FaceKit by the inter-pupillary distance normalized by bizygomatic width, calculated as the ratio of the distance between the two iris centers to the distance between the two zygomatic points (Fig.~\ref{fig:ipd}\subref{fig:ipd-crouzon},\subref{fig:ipd-costello}). This example demonstrates how facial landmarks are converted into an interpretable geometric measurement that can be evaluated at both the individual-patient and disease-cohort levels. Across the 238 patients annotated with the term,
this ratio was elevated by 1.04 reference standard deviations
($q=4.2\times10^{-22}$), and among all 120 measurements it was the most
deviant in these same patients, indicating that the signal is specific to the
measurement the term is mapped to rather than a general distortion of the face.
The same ordering is visible at the cohort level: mean normalized inter-pupillary
distance was 0.532 in the Crouzon syndrome cohort against 0.465 across the
remaining cohorts, the highest of the 50 diseases (Fig.~\ref{fig:ipd}\subref{fig:ipd-bars}). Panels \subref{fig:ipd-crouzon} and \subref{fig:ipd-costello} contrast a synthetic face generated for the Crouzon cohort with one generated for a cohort at the center of the same
distribution, with the two landmark pairs entering the ratio drawn on the image.

For these six terms, the corresponding measurements also deviated in the same
direction when all patients were considered together. The unexpected direction may
therefore partly reflect differences between the patient population and the healthy
reference, rather than an effect specific to patients carrying each annotation.

Several observations suggest possible explanations. For facial asymmetry, the
controls retained more head rotation after filtering, which can make a face appear
less symmetric; patient and control measurements were much closer when compared
within similar head-pose ranges. For full cheeks, the measurement divides cheek
area by squared face width. The annotated patients also had broader, rounder faces,
so a larger denominator may reduce the ratio even when the cheeks appear fuller.
For synophrys, the mapped measurement is the gap between the inner brow endpoints,
whereas the clinical feature is hair occupying that gap, which a geometric distance
does not directly capture.

The remaining cases also warrant caution: the ptosis measurement showed only a
small deviation, the narrow-forehead result was marginally significant and sensitive
to head pose, and the narrow-nasal-bridge result involved few patients and landmark
indices requiring verification. These observations do not establish the cause of
each reversal. They highlight the need to interpret deviations against both healthy
controls and the broader patient population, and to check whether a measurement
and its normalization capture the intended phenotype.

\subsection{Depth for pose correction is better taken from a canonical face than
from the image}
\label{sec:posecorr}

To evaluate the choice of depth used for pose correction
(Equation~\eqref{eq:derot}), we compared four alternatives while keeping all 120
geometric feature formulas unchanged. Method~A applies only a roll alignment from
the eye landmarks, which serves as a baseline. Method~B sets $Z = 0$, treating the face as flat and using the
pose angles alone. Method~C uses the canonical-face table. Method~D uses the per-image
$z$ returned by the landmark detector, scaled by a factor previously tuned
against intra-patient reliability.

\begin{figure}[!htbp]
  \centering
  \begin{subfigure}[b]{0.44\linewidth}
    \centering
    \includegraphics[width=\linewidth]{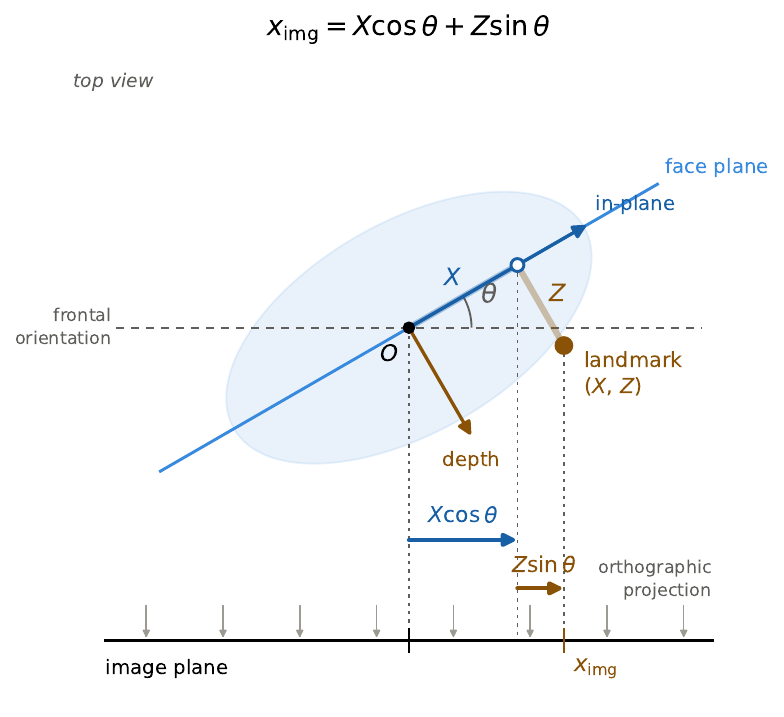}
    \caption{}
    \label{fig:posecorr-geom}
  \end{subfigure}
  \hfill
  \begin{subfigure}[b]{0.54\linewidth}
    \centering
    \includegraphics[width=\linewidth]{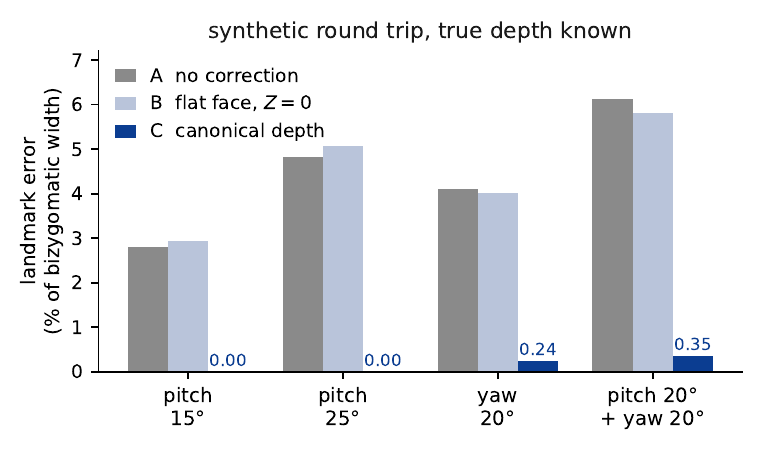}
    \caption{}
    \label{fig:posecorr-roundtrip}
  \end{subfigure}

  \caption{\textbf{Why the depth term is the whole problem.}
  (\protect\subref{fig:posecorr-geom}) Orthographic geometry of a rotated face,
  viewed from above. A landmark lies at $X$ along the face plane and $Z$ along
  the depth direction. When the head turns by $\theta$, its horizontal position
  in the image is the sum of: $X\cos\theta$, which shrinks the
  face horizontally, and $Z\sin\theta$, which pushes the landmark sideways in
  proportion to how deep it sits.
  (\protect\subref{fig:posecorr-roundtrip}) Synthetic round trip with the true
  depth known exactly, giving the residual landmark error as a percentage of
  bizygomatic width for the uncorrected, flat-face and canonical-depth methods at
  four head poses. The flat-face method does not improve on leaving the pose
  uncorrected.}
  \label{fig:posecorr-arms}
\end{figure}

On synthetic faces, where the true depth is known exactly and the
reconstruction can be scored directly, the canonical table recovers the frontal
configuration with little residual error, while the flat-face assumption provides
little improvement over leaving the pose uncorrected: at a pitch of $25^{\circ}$ the residual
landmark error is 4.82\% of bizygomatic width uncorrected, 5.06\% under the
flat-face assumption and 0.00\% with the canonical table, and under a combined
pitch and yaw of $20^{\circ}$ the three are 6.12\%, 5.81\% and 0.35\%
(Fig.~\ref{fig:posecorr-arms}\subref{fig:posecorr-roundtrip}).

On the 4,559 real images the four methods were scored on the subset of measurements
for which independent validity evidence exists, so that the comparison is not
diluted by measurements never shown to measure anything. Two of the three
endpoints require no annotation at all: the absolute correlation between a
measurement and the head-pose angle, which a correct normalization should reduce,
and the intra-patient intraclass correlation across repeat photographs of the
same patient, which sets a ceiling on any downstream model. The third is the area
under the curve for separating patients annotated with a term from those who are
not, on the term-measurement pairs that survived both validity contrasts
(Table~\ref{tab:posearms}).

\begin{table}[!htbp]
  \centering
  \caption{\textbf{Four pose-correction methods on 4,559 GMDB images.}
  Correlations with pitch and yaw and the intraclass correlation (ICC) are means
  over the 13 measurements with independent validity evidence; AUC is the mean
  over the six term--measurement pairs that passed both validity contrasts.
  Lower is better for the pose correlations, higher for ICC and AUC.}
  \label{tab:posearms}
  \begin{tabular}{llccc c}
    \toprule
    Method & Depth $Z$ & $|r|$ pitch & $|r|$ yaw & ICC & AUC \\
    \midrule
    A & none (roll only)      & 0.425 & 0.011 & 0.370 & 0.703 \\
    B & $0$ (flat face)       & 0.422 & 0.010 & 0.377 & 0.703 \\
    C & canonical-face table  & 0.303 & 0.011 & \textbf{0.391} & \textbf{0.714} \\
    D & per-image $z$         & \textbf{0.280} & 0.023 & 0.322 & \textbf{0.714} \\
    \bottomrule
  \end{tabular}
\end{table}

The canonical table reduced the mean
absolute correlation with pitch from 0.425 to 0.303, and raised both the
intra-patient reliability and the discriminative performance, with ICC increasing
from 0.370 to 0.391 and AUC from 0.703 to 0.714. The flat-face method produced only
small changes in pose correlations and ICC, with essentially unchanged AUC.

Using per-image depth reduced the correlation with pitch further, to 0.280, but
intra-patient reliability fell to 0.322, below the 0.370 obtained with roll-only
alignment. Notable losses occurred in
\texttt{nose\_length} (0.413 to 0.269), \texttt{gaze\_asym\_x} (0.469 to 0.327),
and \texttt{forehead\_height} (0.268 to 0.168). The correlation with yaw also increased,
from 0.011 to 0.023, while AUC remained comparable to the canonical-depth method.
These results suggest that per-image depth introduces variability across repeat
photographs without a corresponding gain in discrimination.

The gain from pose correction is concentrated rather than diffuse. Of the six
validated term--measurement pairs, four moved by less than 0.01 in AUC, and the
mean improvement of 0.011 comes almost entirely from two: micrognathia measured
by chin height rose from 0.516 to 0.555, and upslanted palpebral fissures
measured by eye-fissure slant from 0.729 to 0.762. Both involve vertical landmark
positions that are affected by pitch. The comparison therefore supports using
canonical depth to improve repeatability and reduce pose dependence in the
evaluated measurements, with gains in discrimination concentrated in a few
term--measurement pairs.

\begin{figure}[!htbp]
  \centering
  \includegraphics[width=0.92\linewidth]{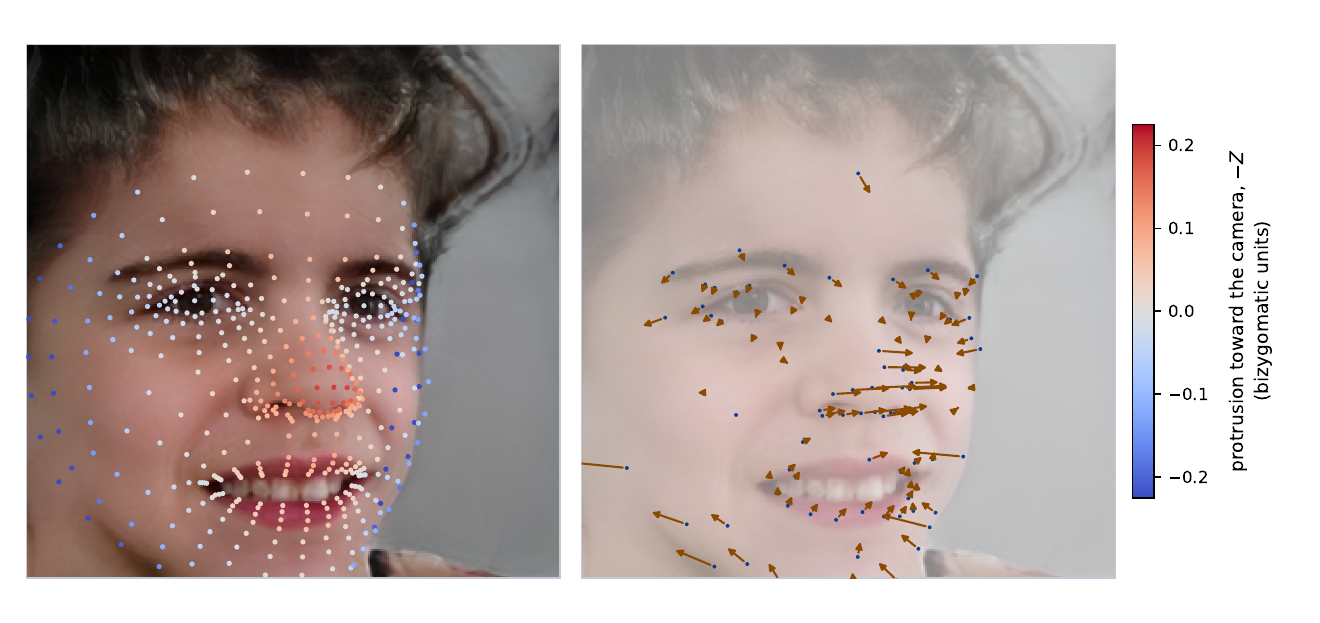}
  \caption{\textbf{What the canonical-depth correction moves on a turned
  face.} A synthetic face generated for the Cornelia de Lange syndrome cohort,
  turned $20^{\circ}$ in yaw; it depicts no real patient. Left: the depth assigned to each landmark, one constant
  per landmark from the canonical table and identical for every image; the
  per-image depth returned by the detector is not read. Right: the displacement
  the correction applies, drawn at twice its true length, with a median of 2.4\%
  of bizygomatic width and a 95th percentile of 7.5\%. The displacement is the
  $Z\sin\theta$ term and is therefore near zero wherever the face is flat and
  largest across the nose and the malar region. It acts on the measurements that
  depend on depth, changing \texttt{face\_asymmetry} by $+53\%$ and
  \texttt{eye\_fissure\_slant\_mean} by $-14\%$ on this image, with a median
  change of 6.1\% over the measurements, and leaves the face-plane ratios alone:
  the inter-pupillary distance over bizygomatic width changes by $-0.08\%$.
  The panels are generated by \texttt{manuscript/make\_posecorr\_mechanism.py},
  which recomputes every number quoted here.}
  \label{fig:posecorr-mech}
\end{figure}

\subsection{The measurements transfer to an independent cohort and degrade
predictably with image resolution}
\label{sec:external}

The results above were obtained on one curated database. Because a measurement
toolkit is only useful if its measurements survive a change of source, we
repeated the central question on RDFace, an independent collection of 456
pediatric facial images spanning 103 rare conditions, released without HPO
annotations and without per-image demographics \cite{rdface}. Only three of its
conditions appear among the 50 GMDB cohorts, and its images are far more
heterogeneous than the GMDB subset, with short sides from 41 to 837 pixels and
roughly a quarter effectively grayscale. No term-level claim can be tested on
this dataset. What it can test is whether the 120 measurements carry disease
information at all on a corpus the pipeline was never tuned against.

FaceKit placed landmarks on 454 of the 456 images. The pose gate passed 348 of
those (76.7\%). Because the RDFace curators had already screened every image as
a frontal portrait under clinical supervision, the 106 rejected images measure
the disagreement between the geometric gate and a human frontality judgement
rather than a share of unusable images, and this disagreement is itself a
sensitivity result: roughly one in four photographs that a clinician accepts as
frontal carries enough residual yaw, pitch or roll to fall outside the
thresholds FaceKit applies.

We then asked whether two images of the same condition are closer in the
120-dimensional measurement space than two images of different conditions,
using the mean within-condition distance minus the mean between-condition
distance as the statistic and restricting the test to conditions with at least
three images. Two confounds were controlled by construction. Images of one
condition are frequently scraped from a single source, so a folder tends to be
homogeneous in resolution and colour; the condition labels were therefore
permuted only within strata defined by image resolution and by colour, which
prevents acquisition similarity from generating the effect. RDFace ships no
patient identifiers, so several images of one condition may show the same
patient; a face-recognition check flagged 10.7\% of within-folder pairs as
falling below an ArcFace threshold calibrated to a 0.1\% false-positive rate on
pairs that cannot be the same person, and those images were removed. A
perceptual difference hash found no repeated image and contributed nothing
beyond the recognition check.

On the 321 images from 81 conditions that pass the gate, images of the same
condition are closer by 1.34 units on the standardized scale, against a
stratified null of $-0.03 \pm 0.12$, placing the observed value 10.7 standard
deviations below the null ($p = 5.0 \times 10^{-4}$, the floor of a
2,000-permutation test). Removing the images flagged as repeat patients leaves
253 images from 67 conditions and a separation 8.2 standard deviations below
the null. The effect therefore does not depend on the same patient appearing
twice.

\begin{figure}[!htbp]
  \centering
  \begin{subfigure}[b]{0.48\linewidth}
    \centering
    \includegraphics[width=\linewidth]{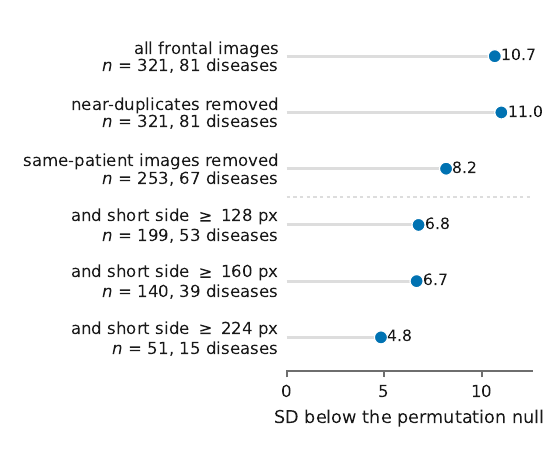}
    \caption{}
    \label{fig:ext-sep}
  \end{subfigure}
  \hfill
  \begin{subfigure}[b]{0.50\linewidth}
    \centering
    \includegraphics[width=\linewidth]{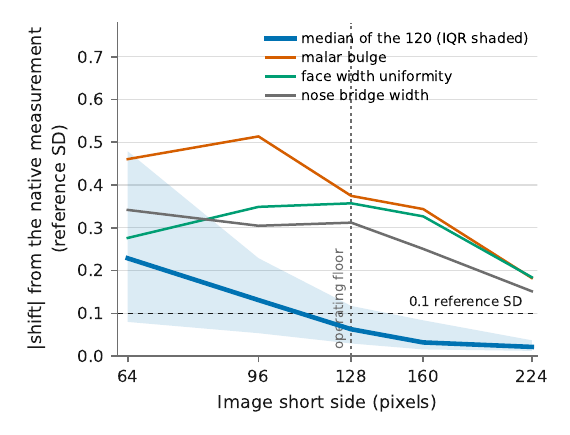}
    \caption{}
    \label{fig:ext-res}
  \end{subfigure}

  \caption{\textbf{The measurements carry disease information outside the
  development database, and the resolution at which they stop being
  comparable.}
  (\protect\subref{fig:ext-sep}) Disease separation in RDFace, expressed as the
  number of standard deviations by which the observed statistic (mean
  within-condition minus mean between-condition distance in the standardized
  120-dimensional space) falls below a permutation null in which the condition
  labels are shuffled only within strata of image resolution and colour, so
  that acquisition similarity cannot produce the effect. The controls are
  cumulative from top to bottom: perceptual near-duplicates removed, then images
  that a face-recognition check assigns to a patient already present, then a
  ladder of resolution gates. Every row reaches the floor of the
  2,000-permutation test ($p = 5.0 \times 10^{-4}$). The separation weakens as
  the gates remove images but does not disappear, which is what distinguishes it
  from an artifact of repeated patients or of the low-resolution tail.
  (\protect\subref{fig:ext-res}) Paired within-image drift: every control image
  is measured at its native 448 pixels and again after downsampling, and the
  change is expressed in reference standard deviations. The heavy line is the
  median over the 120 measurements with the interquartile range shaded; the
  three thin lines are the measurements displaced most at 128 pixels. The
  catalogue is stable to about 128 pixels, where the median shift is 0.06
  reference SD and 36 of the 120 measurements exceed 0.1, and degrades quickly
  below it. Panels are generated by
  \texttt{manuscript/make\_sensitivity\_panels.py}.}
  \label{fig:external}
\end{figure}

Because RDFace reaches much lower resolutions than the GMDB subset, we repeated
the test behind a ladder of resolution gates. The separation persists at every
level: $6.8$ standard deviations below the null at a short side of at least 128
pixels (199 images), $6.7$ at 160 pixels (140 images), and $4.8$ at 224 pixels,
the resolution at which the GMDB patients are measured, where only 51 images
and 15 conditions remain. The measurement space therefore carries disease
information outside the database FaceKit was developed on, and the signal is not
produced by the unreliable low-resolution tail
(Fig.~\ref{fig:external}\subref{fig:ext-sep}).

The resolution dependence itself was quantified directly, as a paired
within-image quantity that involves no model: every control image was measured
at its native 448 pixels and again after downsampling, and the change in each
measurement was expressed in reference standard deviations. The catalogue is
stable down to about 128 pixels and degrades quickly below it. The median
absolute shift across the 120 measurements is 0.02 reference SD at 224 pixels,
0.06 at 128 pixels, 0.13 at 96 pixels and 0.23 at 64 pixels, and the number of
measurements displaced by more than 0.1 reference SD rises from 9 to 36 to 67
to 86 over the same range. The measurements most affected are those defined on
low-contrast contours, led by \texttt{malar\_bulge\_mean},
\texttt{face\_width\_uniformity} and \texttt{nose\_bridge\_width}
(Fig.~\ref{fig:external}\subref{fig:ext-res}). We therefore
report 128 pixels as the operating floor of the catalogue, and images below it
should be treated as carrying a systematic offset rather than only added noise.

Finally, for the three conditions shared with GMDB we compared the mean
measurement profile of the RDFace images against all 50 GMDB cohort profiles.
The matching GMDB cohort ranked third, third and sixth of 50 after centering
each profile on the cohort mean (Angelman syndrome $r = 0.40$, Joubert syndrome
$r = 0.38$, Mowat--Wilson syndrome $r = 0.37$), and fourth, fourth and fifth
without centering. The matching cohort is thus consistently among the closest
few of 50 but is never the single closest, which is the expected outcome given
that each RDFace condition contributes only three to five images.

\subsection{Age and image resolution change the measurements but not the
reference}
\label{sec:age}

The reference distributed with FaceKit pools control images below the age of
20, while the median age of the GMDB cohorts ranges from 1.1 to 15.0 years. Two
separate questions follow: whether the measurements depend on age, and whether
conditioning the reference on age changes any conclusion. They have different
answers.

The measurements do depend on age. In the healthy controls, with ancestry,
gender and the three pose angles adjusted for, the partial $\eta^2$ of age band
has a median of 0.038 across the 120 measurements and reaches 0.28 for
\texttt{nose\_tip\_midline\_dent}, 0.25 for \texttt{chin\_height} and 0.24 for
\texttt{face\_aspect\_ratio}. Facial proportion changes with growth, and the
catalogue records that change.

Conditioning the reference on it nonetheless buys nothing. We re-scored the
same patients against a reference restricted to their own age band, changing
only the reference: the patients, the measurements, the statistics and the
thresholds are those of the analysis in Section~\ref{sec:validity}, and the two
arms are compared term by term on the same patients. Of the 77 terms testable
in the 148 patients with a usable recorded age, 24 were reproduced against the
pooled reference and 27 against the age-conditioned one, four terms gaining and
one losing (exact McNemar $p = 0.375$). Across all 77 terms the signed effect
moved by a median of $-0.0006$ (Wilcoxon $p = 0.945$), the mean absolute effect
went from 0.472 to 0.480, and the number of terms whose mapped measurement
ranked in the top 30 was unchanged at 20.

\begin{figure}[!htbp]
  \centering
  \begin{subfigure}[b]{0.325\linewidth}
    \centering
    \includegraphics[width=\linewidth]{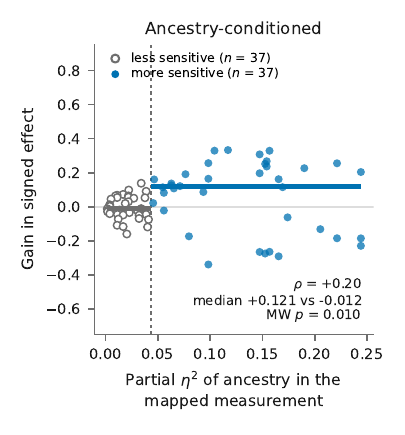}
    \caption{}
    \label{fig:ref-anc}
  \end{subfigure}
  \hfill
  \begin{subfigure}[b]{0.325\linewidth}
    \centering
    \includegraphics[width=\linewidth]{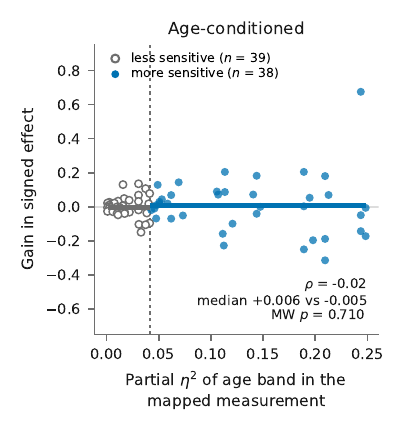}
    \caption{}
    \label{fig:ref-age}
  \end{subfigure}
  \hfill
  \begin{subfigure}[b]{0.325\linewidth}
    \centering
    \includegraphics[width=\linewidth]{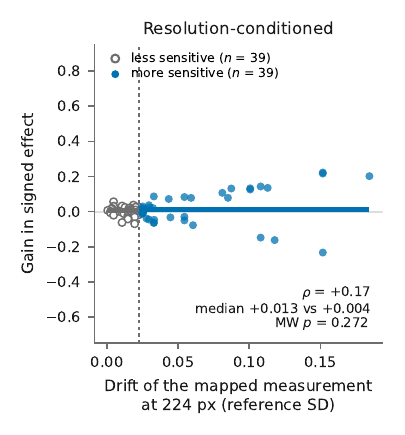}
    \caption{}
    \label{fig:ref-res}
  \end{subfigure}

  \caption{\textbf{Of the three variables on which the distributed reference
  differs from the patients, only ancestry changes what the reference
  concludes.}
  Each point is one HPO term, and the three panels differ only in the variable
  the reference is conditioned on: ancestry (\protect\subref{fig:ref-anc}), age
  band (\protect\subref{fig:ref-age}) and image resolution
  (\protect\subref{fig:ref-res}). The abscissa is the sensitivity of the
  measurement that term is mapped to, measured in the healthy controls: for the
  first two, the partial $\eta^2$ of that variable, each adjusted for the other,
  for gender and for the three pose angles; for the third, the median shift of
  the measurement when the same image is measured at 224 rather than 448 pixels,
  a paired quantity that involves no model. The ordinate is the change in the
  term's signed effect when the same patients are re-scored against the
  conditioned reference instead of the pooled one; everything except the
  reference is held fixed, and all three panels share the ordinate scale. The
  dashed line splits the terms at the median sensitivity and the horizontal bars
  are the median gain in each half. A conditioning that removes a real confound
  should lift the right-hand half and leave the left-hand half alone, which is
  what ancestry does and what neither age nor resolution does. This contrast,
  rather than the number of terms that cross a significance threshold, is what
  separates the three: the headline counts move by six, four and one term
  respectively, none of which is significant on a catalogue of this size. Panels
  are generated by \texttt{manuscript/make\_sensitivity\_panels.py}.}
  \label{fig:refcond}
\end{figure}

The reading that settles it is not the size of the change but where it falls. A
conditioning that removes a real confound should help most on the measurements
that are most sensitive to what is being conditioned on. It does not: the
correlation between the age sensitivity of a term's mapped measurement and the
gain that term receives is $\rho = -0.02$ ($p = 0.881$), and splitting the terms
at the median of that sensitivity gives median gains of $+0.006$ and $-0.005$ in
the two halves (Mann--Whitney $p = 0.710$; Fig.~\ref{fig:refcond}\subref{fig:ref-age}).
The small movement that does occur is indistinguishable from drift. Conditioning on any variable also shrinks the
reference standard deviation and inflates every $z$ mechanically; here the
within-band to pooled SD ratio is 0.985, bounding that artifact at about 1.5\%,
which is of the same order as the entire observed change. We therefore ship a
single age-pooled reference, and note that this conclusion is specific to a
reference already restricted to individuals under 20.

Image resolution, the third variable on which the reference and the patients
differ, behaves the same way. The distributed reference is built from 448-pixel
control images while the patients are measured at 224 pixels, and the previous
subsection showed that this difference is not free: the median measurement moves
by 0.02 reference SD between the two resolutions and nine measurements move by
more than 0.1. We therefore rebuilt the reference from the same control images
re-measured at 224 pixels and re-scored the patients against it. Because this
arm re-measures the controls rather than partitioning them, the reference
standard deviation does not shrink and the mechanical inflation that conditioning
on a grouping variable produces cannot arise here at all. Of the 78 terms
testable in 238 patients, one changed verdict (\emph{Square face}, mapped to
\texttt{face\_width\_uniformity}, $q$ from 0.068 to 0.020), taking the
reproduced count from 33 to 34 (exact McNemar $p = 1.0$); the signed effect moved
by a median of $+0.005$ (Wilcoxon $p = 0.203$), the mean absolute effect from
0.465 to 0.487, and the number of terms whose mapped measurement ranked in the
top 30 was unchanged at 26. As with age, the gain does not concentrate where it
should: $\rho = +0.17$ between a term's resolution sensitivity and its gain
($p = 0.134$), with median gains of $+0.013$ and $+0.004$ across the median split
(Mann--Whitney $p = 0.272$; Fig.~\ref{fig:refcond}\subref{fig:ref-res}).
Resolution is worth controlling at acquisition, as Section~\ref{sec:external}
shows, but it is not worth conditioning the reference on.

Of the three variables on which the distributed reference differs from the
patient population, only one changes a conclusion, and it is the subject of the
next subsection.

\subsection{Ancestry enters through the reference, not through the disease
effect}
\label{sec:ancestry}

Ancestry behaves in the opposite way to the other two, and the distinction
matters for how a FaceKit $z$-score should be read.

The measurements are ancestry-sensitive to a similar degree as they are
age-sensitive, with a median partial $\eta^2$ of 0.032 in the controls after
adjustment for age band, gender and pose, rising to 0.24 for
\texttt{eye\_fissure\_length\_mean}, 0.22 for
\texttt{inter\_pupillary\_distance} and 0.21 for
\texttt{canthal\_\allowbreak to\_\allowbreak pupillary\_\allowbreak ratio}. Unlike age and resolution, however,
conditioning the reference on ancestry moves the conclusions in the direction a
real confound predicts. Of the 74 terms testable in the 194 patients with a recorded
ancestry, 29 were reproduced against the pooled reference and 34 against the
ancestry-conditioned one, six terms gaining and one losing (exact McNemar
$p = 0.125$), and the mean absolute effect rose from 0.474 to 0.536. The
headline count is underpowered on a catalogue of this size, but the gain
concentrates where it should: the correlation between the ancestry sensitivity
of a term's mapped measurement and the gain that term receives is
$\rho = +0.20$, and the half of the terms whose measurements are more
ancestry-sensitive gains a median of $+0.121$ against $-0.012$ for the less
sensitive half (Mann--Whitney $p = 0.0098$;
Fig.~\ref{fig:refcond}\subref{fig:ref-anc}). The mechanical inflation from a
narrower reference is bounded at 2.0\% by the within-group to pooled SD ratio of
0.980, an order of magnitude smaller than the gain in the sensitive half. Where
ancestry is recorded, a conditioned reference is therefore the better choice,
and \texttt{facekit score} accepts one.

That the reference depends on ancestry does not imply that a disease effect
does, and the two are separable in a cohort where both a case group and a
control group are available in every ancestry group. We measured 457 clinical
photographs of patients with 22q11.2 deletion syndrome, of which 434 carry a
recorded ancestry (352 European, 51 African, 31 Asian), and compared them
against the control images that pass a strict pose filter of $8^{\circ}$ on all
three angles (176, 95 and 87 respectively). For each of the 120 measurements we
fitted
\[
  \text{measurement} \sim C(\text{ancestry}) \times C(\text{cohort})
    + \text{yaw} + \text{pitch} + \text{roll}
\]
and read the ancestry-by-cohort interaction from a type-II analysis of
variance, with Benjamini--Hochberg correction across the 120 measurements. The
cohort main effect is deliberately not interpreted: the patients are clinical
photographs and the controls are in-the-wild images, so a site and acquisition
difference is confounded with the disease. The interaction is the quantity that
is free of it, and it asks the question that matters for a distributed
reference, namely whether the deviation a disease produces is the same size in
every ancestry group.

\begin{figure}[!htbp]
  \centering
  \begin{subfigure}[b]{0.51\linewidth}
    \centering
    \includegraphics[width=\linewidth]{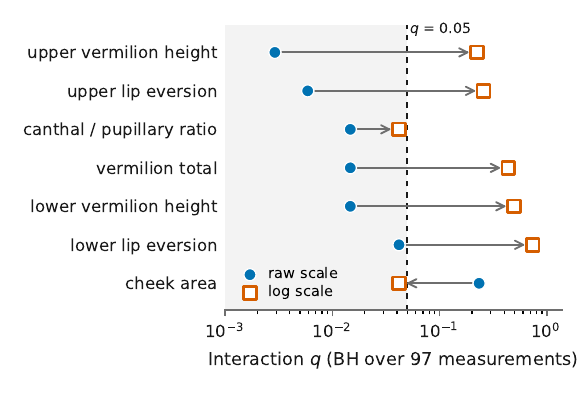}
    \caption{}
    \label{fig:anc-scale}
  \end{subfigure}
  \hfill
  \begin{subfigure}[b]{0.47\linewidth}
    \centering
    \includegraphics[width=\linewidth]{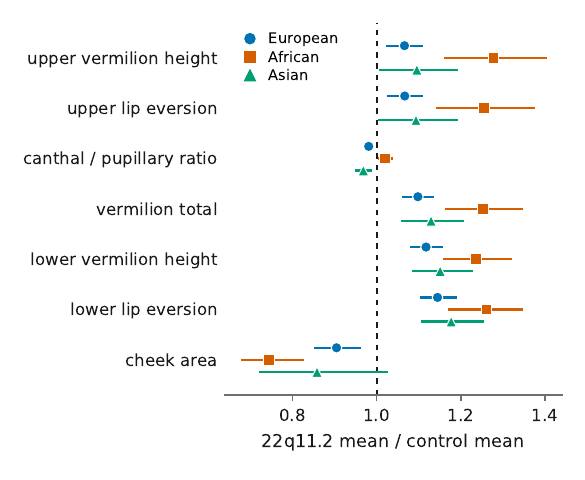}
    \caption{}
    \label{fig:anc-ratio}
  \end{subfigure}

  \caption{\textbf{The 22q11.2 deletion phenotype deviates from controls by the
  same proportion in every ancestry group.}
  (\protect\subref{fig:anc-scale}) Ancestry-by-cohort interaction $q$ for every
  measurement significant on at least one scale, fitted on the raw measurement
  and on its logarithm with the identical model and the identical rows. The
  correction family is the 97 strictly positive measurements, so the two arms
  are directly comparable. The four lip measurements cross from below to above
  $q = 0.05$ once the scale is changed, which is what an effect that is
  multiplicative rather than additive does;
  \texttt{canthal\_\allowbreak to\_\allowbreak pupillary\_\allowbreak ratio} barely moves and is the only
  measurement significant on both scales.
  (\protect\subref{fig:anc-ratio}) The same measurements as a ratio of the
  22q11.2 mean to the control mean within each ancestry group, with 95\%
  bootstrap intervals over 2,000 resamples. The lip measurements sit above 1 in
  all three groups and differ only in how far above, on a baseline that is
  itself largest in the African controls; only
  \texttt{canthal\_\allowbreak to\_\allowbreak pupillary\_\allowbreak ratio} crosses 1, changing direction between
  groups. Panels are generated by
  \texttt{manuscript/make\_sensitivity\_panels.py}.}
  \label{fig:anc}
\end{figure}

Five of the 120 measurements reached $q < 0.05$: four lip measurements and
\texttt{canthal\_\allowbreak to\_\allowbreak pupillary\_\allowbreak ratio}. Four of the five do not survive a change
of scale. Every lip deviation has the same sign in all three ancestry groups and
differs only in magnitude, with a case-to-control ratio of 1.07 to 1.10 in the
European and Asian groups against 1.24 to 1.28 in the African group, on a
baseline that is itself largest in the African controls
(Fig.~\ref{fig:anc}\subref{fig:anc-ratio}). This is the signature of a
multiplicative effect read on an additive scale. We therefore refitted the
identical model on the logarithm of each measurement. The scale comparison is
restricted to the 97 measurements that are strictly positive and therefore
testable on both scales, so that the raw and the log arm share one correction
family. Six of those 97 reach $q < 0.05$ on the raw scale and two on the log
scale. The four lip terms fall from $q = 0.0029$ to $0.042$ down to $q = 0.22$ to
$0.74$, while \texttt{canthal\_\allowbreak to\_\allowbreak pupillary\_\allowbreak ratio} moves only from $0.015$ to
$0.042$ and is the one measurement significant on both scales, with a partial
$\eta^2$ of 0.019 (Fig.~\ref{fig:anc}\subref{fig:anc-scale}). A second
measurement, \texttt{cheek\_area\_mean}, is
significant on the log scale ($q = 0.042$) and not on the raw scale
($q = 0.23$); with one such term out of 97 we do not interpret it.

Two acquisition artifacts were excluded rather than assumed away. Mouth opening
does not drive the lip result: no image in either group exceeds the aperture
threshold at which FaceKit masks 

the lip measurements, and the aperture
correlates with the four lip measurements at $|r| \le 0.14$. Resolution cannot
generate the interaction: all 457 patient images are 784 pixels and all control
images 448 pixels, so resolution differs between the groups but is constant
within each and identical across ancestry, loading entirely on the cohort main
effect that this design already declines to interpret. What remains uncontrolled
is skin tone interacting with the placement of the vermilion border under
different acquisition conditions, which would appear in this design as an
ancestry-by-cohort interaction and is the reason the log-scale result is
reported alongside the raw one rather than in place of it.

Taken together, the two analyses locate the ancestry dependence in the baseline
rather than in the disease signal. A FaceKit $z$-score compares a patient to a
reference population whose composition matters, so the reference should be
conditioned on ancestry when it is recorded; but the deviation a syndrome
produces was, for this syndrome and these 120 measurements, proportionally the
same in all three ancestry groups. A practical consequence follows for ratio
measurements, which most of the catalogue consists of: group comparisons on
them are better made on the log scale, where a proportional effect is additive
and a difference in baseline does not masquerade as an interaction.

\subsection{Synthetic images reproduce cohort-level geometric phenotypes}

We next asked whether the synthetic faces produced by the released generator
carry the same geometric phenotype signal as the real photographs it was trained
to imitate. For the 50 rare-disease cohorts we compared 4,559 synthetic images,
sampled so that every cohort contributes as many synthetic images as it has real
ones, against the 4,559 real training images, passing both sets through the
identical FaceKit pipeline and comparing them on the same 120 measurements. For
each syndrome the comparison was restricted to the measurements its HPO
annotations map to. One cohort, intellectual disability, autosomal dominant 50,
has no mapped measurement and enters the analysis only as part of the comparison
pool, which leaves 1,013 syndrome--measurement pairs across 49 cohorts.

\begin{figure}[!htbp]
  \centering
  \begin{subfigure}[b]{0.5\linewidth}
    \centering
    \includegraphics[width=\linewidth,height=0.23\textheight,keepaspectratio]{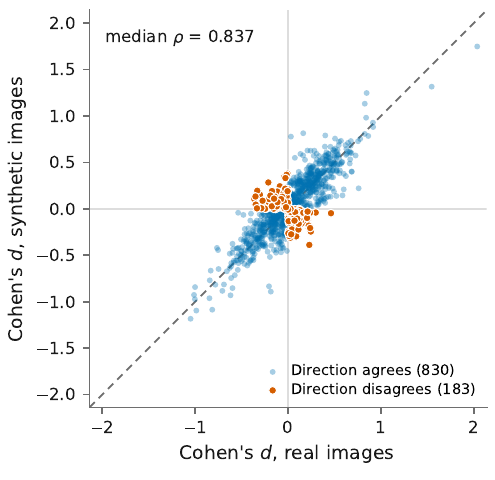}
    \caption{}
    \label{fig:syn-scatter}
  \end{subfigure}
  \hfill
  \begin{subfigure}[b]{0.47\linewidth}
    \centering
    \includegraphics[width=\linewidth,height=0.23\textheight,keepaspectratio]{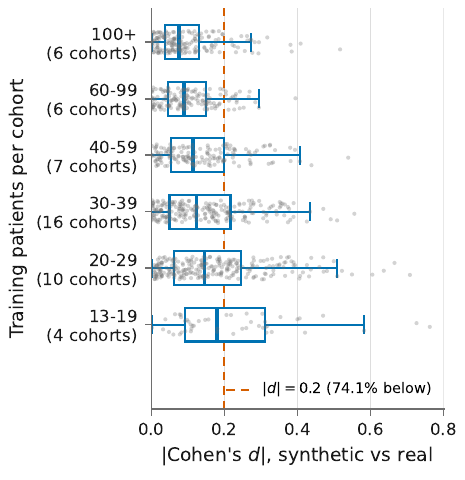}
    \caption{}
    \label{fig:syn-within}
  \end{subfigure}

  \vspace{0.4em}

  \begin{subfigure}[b]{\linewidth}
    \centering
    \includegraphics[width=\linewidth,height=0.18\textheight,keepaspectratio]{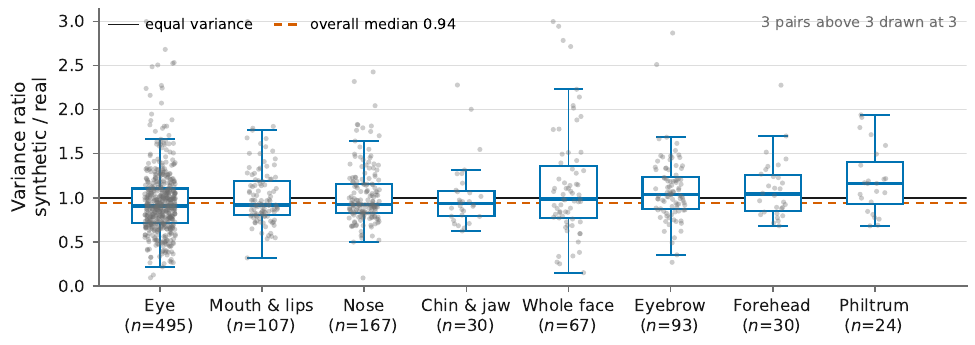}
    \caption{}
    \label{fig:syn-var}
  \end{subfigure}

  \caption{\textbf{Synthetic cohorts reproduce the geometric phenotype of the
  cohorts they were trained on.} All panels cover the 1,013 syndrome--measurement
  pairs obtained by restricting each of the 49 mapped syndromes to the
  measurements its HPO annotations map to. (\protect\subref{fig:syn-scatter})
  Effect of the focal cohort against the remaining cohorts pooled, measured in
  the synthetic images against the same effect measured in the real images; the
  dashed line is equality. Pairs whose effect disagrees in sign between the two
  are highlighted, and lie almost entirely near the origin, where the real effect
  is itself close to zero. The median $\rho$ is taken over cohorts with at least
  three measurements. (\protect\subref{fig:syn-within}) Within each syndrome, the
  absolute Cohen's $d$ between the synthetic and the real distribution of a
  measurement, grouped by the number of training patients per cohort. Boxes give
  the median and interquartile range, whiskers extend to $1.5\times$ the
  interquartile range, and individual pairs are drawn behind them.
  (\protect\subref{fig:syn-var}) Ratio of the variance of a measurement in the
  synthetic images to its variance in the real images, by facial region and
  ordered by median. Values below one indicate a synthetic cohort narrower than
  the real cohort; three ratios above 3, each caused by a single synthetic image
  with an outlying landmark, are drawn at 3.}
  \label{fig:syn}
\end{figure}

Within each syndrome, the synthetic and real distributions of a measurement were
close. The overall median absolute Cohen's $d$ between the two sets was 0.111;
74.1\% of all pairs fell below $|d| = 0.2$ and 98.6\% below $|d| = 0.5$
(Fig.~\ref{fig:syn}\subref{fig:syn-within}). Among the 39 cohorts mapped to at
least ten measurements, the median ranged from 0.041
(intellectual disability-sparse hair-brachydactyly syndrome) to 0.210 (Joubert
syndrome). The median was smallest in the cohorts with the most training patients
(0.076 for cohorts with 100 or more) and largest in those with the fewest (0.181
for cohorts with 13 to 19), which also contribute the fewest images to both sets.
A generator therefore reproduces not only the appearance of a cohort but the
location and spread of the individual measurements taken from it.

A more demanding test is whether the synthetic images preserve what
distinguishes one syndrome from the others, rather than only reproducing each
cohort in isolation. For every syndrome we computed the effect of the focal
cohort against the 49 remaining cohorts pooled, separately in the synthetic
and in the real images, and compared the two sets of effects. The direction of
the effect agreed between synthetic and real for a median of 86.7\% of a
syndrome's measurements, and the effect sizes themselves tracked closely
(median Spearman $\rho = 0.837$, median Pearson $r = 0.866$;
Fig.~\ref{fig:syn}\subref{fig:syn-scatter}). Among the cohorts mapped to at least
ten measurements, agreement was weakest for Costello syndrome (57.7\% of 26
measurements), Weaver syndrome (62.5\% of 16) and Menke-Hennekam syndrome (65.2\%
of 46), and complete for Floating-Harbor syndrome (16 measurements). Agreement
declined with the number of training patients: the median Spearman $\rho$ was
0.945 for cohorts with 60 to 99 patients and 0.835 for those with 100 or more,
against 0.771 for cohorts with 20 to 29 and 0.604 for the four cohorts with 13
to 19.

Across all 1,013 pairs, 183 (18.1\%) disagreed in sign
(Fig.~\ref{fig:syn}\subref{fig:syn-scatter}, highlighted). Disagreement was most
frequent among the whole-face measurements (19 of 67 pairs, 28.4\%) and in the eye
region (100 of 495, 20.2\%), and least frequent in the philtrum (2 of 24, 8.3\%).
The disagreements were largely confined to measurements that barely separate the
cohorts in the real images to begin with: the median $|d|$ measured in real images
was 0.077 for the disagreeing pairs against 0.237 for the agreeing ones, and 83.6\%
of the disagreements occurred where the real effect was below $|d| = 0.2$. Where a
measurement carries no discriminative signal in the real cohort, its sign is not
determined, so these cases indicate an absence of signal to reproduce rather
than a signal reproduced incorrectly. The largest real effects that changed sign
concerned palpebral fissure slant, in Au-Kline syndrome ($d = -0.048$ synthetic
against $d = +0.463$ real), Cohen syndrome and autosomal recessive Robinow
syndrome ($|d| = 0.36$ and $0.35$ in real images).

The one systematic difference between synthetic and real images is dispersion,
and it is modest. The variance of a measurement in the synthetic set was smaller
than in the real set for 58.1\% of pairs, with a median ratio of 0.94. The
narrowing was concentrated in the eye (median ratio 0.91), mouth and lips (0.92),
nose (0.92) and chin and jaw (0.94), whereas the eyebrow and forehead measurements
(1.04 each) and the philtrum (1.16) were not narrower than in the real images
(Fig.~\ref{fig:syn}\subref{fig:syn-var}). The generator therefore recovers where a
syndrome sits in measurement space and how it differs from other syndromes, and
slightly under-represents the variation of the central facial features within it.
This should be taken into account when synthetic images are used to augment
training data.

\begin{table}[!htbp]
  \centering
  \caption{\textbf{Synthetic images flagged against a real-image reference
  rate.} Each row fixes an operating point by taking the threshold from the
  $p$-th percentile of the distances between held-out real images and the
  training partition, so that $p$ is the false acceptance rate the threshold
  produces among real images of the same cohort; entries give the percentage of
  the 4,746 synthetic images with a detected face (4,785 for LPIPS) falling
  below that threshold. Thresholds are cosine distances for the three
  recognition models and LPIPS distances for the perceptual metric. Values at
  or below $p$ indicate no excess over the reference.}
  \label{tab:privacy-far}
  \begin{tabular}{cccccccc}
    \toprule
    & \multicolumn{3}{c}{Identity (\% flagged)} & \multicolumn{3}{c}{Threshold} & Appearance \\
    \cmidrule(lr){2-4} \cmidrule(lr){5-7} \cmidrule(lr){8-8}
    $p$ (\%) & ArcFace & AdaFace & LVFace & ArcFace & AdaFace & LVFace & LPIPS (\% flagged) \\
    \midrule
     1 &  0.17 &  0.13 &  0.15 & 0.3239 & 0.3115 & 0.5046 &  1.0 \\
     2 &  0.65 &  0.78 &  0.65 & 0.3860 & 0.3696 & 0.5454 &  2.7 \\
     5 &  2.78 &  3.39 &  3.01 & 0.4525 & 0.4248 & 0.6076 &  8.6 \\
    10 &  6.09 &  7.42 &  6.01 & 0.4977 & 0.4641 & 0.6427 & 15.5 \\
    20 & 12.73 & 16.01 & 15.40 & 0.5423 & 0.5171 & 0.6936 & 26.4 \\
    \bottomrule
  \end{tabular}
\end{table}

\subsection{Synthetic images show no detectable excess similarity to training
identities}

\begin{figure}[!htbp]
  \centering
  \begin{subfigure}[b]{0.48\linewidth}
    \centering
    \includegraphics[width=\linewidth]{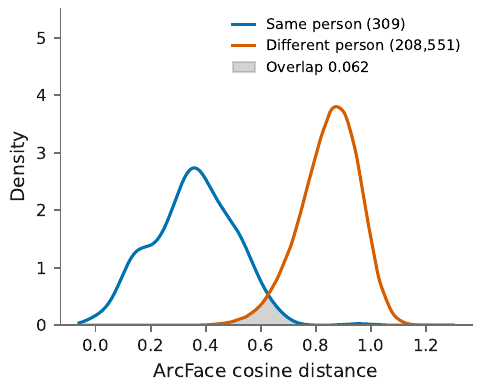}
    \caption{}
    \label{fig:privacy-kde-arcface}
  \end{subfigure}
  \hfill
  \begin{subfigure}[b]{0.48\linewidth}
    \centering
    \includegraphics[width=\linewidth]{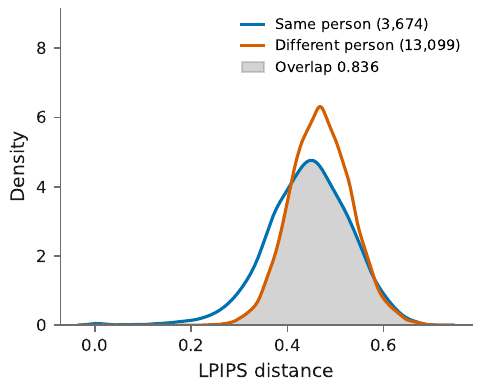}
    \caption{}
    \label{fig:privacy-kde-lpips}
  \end{subfigure}

  \vspace{0.4em}

  \begin{subfigure}[b]{0.48\linewidth}
    \centering
    \includegraphics[width=\linewidth]{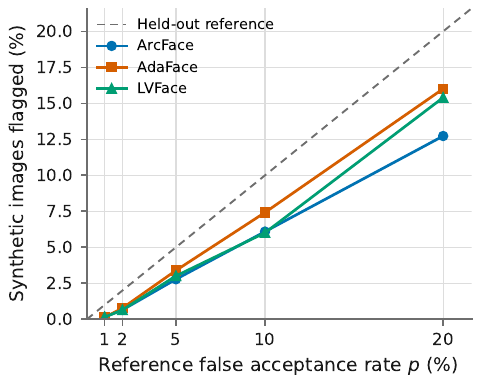}
    \caption{}
    \label{fig:privacy-far}
  \end{subfigure}
  \hfill
  \begin{subfigure}[b]{0.48\linewidth}
    \centering
    \includegraphics[width=\linewidth]{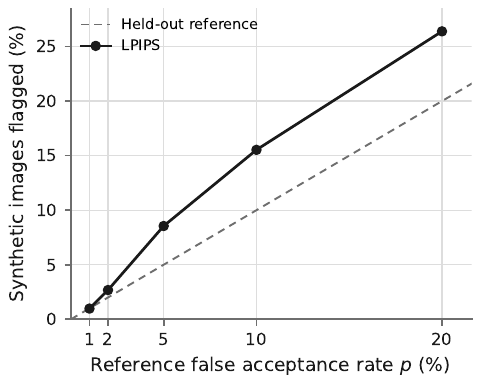}
    \caption{}
    \label{fig:privacy-far-lpips}
  \end{subfigure}

  \caption{\textbf{Privacy assessment of the audited generator.}
  (\protect\subref{fig:privacy-kde-arcface}) Distribution of ArcFace cosine
  distances between images of one patient and between images of different
  patients, with the overlap between the two densities shaded.
  (\protect\subref{fig:privacy-kde-lpips}) The same distributions measured with
  LPIPS, which overlap almost completely and therefore carry little identity
  information. Panels (\protect\subref{fig:privacy-kde-arcface}) and
  (\protect\subref{fig:privacy-kde-lpips}) rest on different numbers of
  same-person pairs, 309 against 3,674, because the half-year restriction on the
  interval between two images of one patient applies to the recognition models
  only. (\protect\subref{fig:privacy-far}) Percentage of synthetic images
  falling below a threshold set at the $p$-th percentile of the held-out
  distances, for the three recognition models; the dashed line is the rate the
  same threshold produces among real held-out images.
  (\protect\subref{fig:privacy-far-lpips}) The same comparison measured with
  LPIPS, where the synthetic images are flagged above the held-out reference at
  every operating point from 2\% upward.}
  \label{fig:privacy}
\end{figure}

Because the generator is distributed with FaceKit, we finally asked whether
the images it produces disclose the identity of the patients whose photographs
trained them. The audit requires real images that the generator has never seen,
so it was carried out on a generator of the same architecture and training
schedule fitted to the training partition alone, 3,602 images of 2,547 patients
from the 50 cohorts; the released generator is trained on all 4,559 images. From
the audited generator we sampled 4,785 synthetic images, five for every held-out
image of each cohort, of which 4,746 contained a detectable face, and compared
them against the training partition, using as a reference the 957 held-out real
images of 661 patients that belong to the same cohorts but were excluded from
generator training. Face recognition embeddings separate these cohorts less
cleanly than they separate adults in the benchmarks they were calibrated on:
across 309 same-person pairs, restricted to images of one patient taken less
than half a year apart, and 208,551 different-person pairs, the two ArcFace
distance distributions were largely distinct but retained a tail overlap of
0.062 (Fig.~\ref{fig:privacy}\subref{fig:privacy-kde-arcface}; 0.059 for AdaFace
and 0.057 for LVFace), so no threshold separates identities without error. We
therefore did not fix a verification threshold. Each threshold was instead read
off the held-out distances themselves, so that an operating point is defined by
the false acceptance rate it produces among real images of the same cohort, and
the synthetic images were asked only to stay at or below that rate. A held-out
image is subject to the same recognition errors and to the same coincidental
resemblance between children who share a syndrome as a synthetic image is, so
what remains above this reference is the excess attributable to the generator.
The comparison requires the embedding to rank identities correctly rather than
to be calibrated, and it does: each same-person pair contributes two anchor
orderings, and in 607 of these 618 comparisons (98.2\%) the distance of the
same-person pair was smaller than every different-person distance it was
compared against with ArcFace, with 98.2\% for AdaFace and 97.9\% for LVFace.

Against this reference the synthetic images carried no excess identity signal
(Table~\ref{tab:privacy-far},
Fig.~\ref{fig:privacy}\subref{fig:privacy-far}). At an operating point
corresponding to a 1\% false acceptance rate among real images, 0.17\% of
synthetic images fell below the ArcFace threshold, 0.13\% below the AdaFace
threshold and 0.15\% below the LVFace threshold. The margin persisted as the
operating point was relaxed: at 5\% the three rates were 2.78\%, 3.39\% and
3.01\%, and at 20\% they were 12.73\%, 16.01\% and 15.40\%. No backbone exceeded
its reference rate at any of the five operating points. Within the resolution of
these three recognition models, and at every operating point examined, sampling
from the generator is therefore not more likely to produce a face matching a
training patient than drawing another real photograph of the same syndrome is.

The cohorts differ more than tenfold in the number of patients available for
training, from 13 to 242, and a generator might be expected to reproduce
individuals more readily where it has seen few of them. For every cohort and
recognition model we therefore computed the probability that a synthetic image
lies closer to the training partition than a held-out image of the same cohort
does, a value of one half indicating no excess. Across the 50 cohorts this
probability showed no association with the number of training patients
(Spearman $\rho$ from $-0.18$ to $-0.10$ for the three models, permutation
$p \geq 0.19$). Pooled within tiers of training-patient count, no tier had a 95\%
bootstrap interval lying above one half. The highest estimates came from the four
cohorts with 13 to 19 training patients, 0.55 for ArcFace, 0.62 for AdaFace and
0.48 for LVFace, whose intervals were wide and included one half in each case.

The picture on the appearance axis is different, and the two must not be
conflated. LPIPS distances between images of one patient and between images of
different patients are almost indistinguishable, overlapping by 0.836 across
3,674 same-person and 13,099 different-person pairs. The same-person count is
larger than the 309 pairs of the recognition analysis because the half-year
restriction is not applied here, LPIPS not being read as an identity metric; the
pairs are otherwise drawn from the same images
(Fig.~\ref{fig:privacy}\subref{fig:privacy-kde-lpips}); the metric responds to
texture, pose and illumination rather than to identity. Measured with it, the
synthetic images were closer to the training photographs than the held-out
images were: 8.6\% of synthetic images fell below the threshold set at a 5\%
reference rate, and 26.4\% below the threshold set at 20\%
(Fig.~\ref{fig:privacy}\subref{fig:privacy-far-lpips}).
A generator trained on a cohort reproduces the acquisition characteristics of
that cohort, and the synthetic images inherit them. Because the same metric
cannot tell two photographs of one patient from photographs of two different
patients, this proximity describes the distribution the images were drawn from
and not the individuals within it.

Both axes are summarized by the nearest neighbor adversarial accuracy, computed
on 1,000 training, 957 held-out and 1,000 synthetic images for the embeddings
and on 200 images per set for LPIPS, with 1,000 bootstrap replicates. Computed on
embeddings, the privacy loss was $+0.013$ for ArcFace ($[-0.000, 0.028]$),
$+0.020$ for AdaFace ($[0.003, 0.036]$) and $+0.012$ for LVFace
($[-0.003, 0.027]$). The intervals for ArcFace and LVFace included zero; the
interval for AdaFace excluded it narrowly. A difference of two percentage points
in adversarial accuracy is small, and it is not accompanied by any excess in the
flagging analysis, where every recognition model stayed below its reference rate
at every operating point; we read it as a slight shift of the synthetic images
toward the training partition as a whole rather than as the reproduction of
individual patients. Computed on LPIPS distances the privacy loss was $+0.038$,
with a confidence interval of $[-0.030, 0.103]$ that is wide on 200 images per
set and includes zero. One feature of these values deserves comment: the
adversarial accuracies themselves ranged from 0.903 to 0.939 on embeddings, far
above the value of one half that would indicate two indistinguishable sets,
because synthetic faces are globally separable from real ones in recognition
embedding space. The absolute level therefore carries no privacy
interpretation here, and only the difference between the two comparisons does.

\FloatBarrier

\section{Discussion}

FaceKit provides a reproducible and interpretable workflow for transforming facial images into structured geometric phenotype representations. The phenotype-faithfulness analysis showed that most of the mapped measurements deviated from a normative reference in the direction encoded by the corresponding HPO term, although only a subset of these deviations reached statistical significance. At the same time, the results demonstrate that the feature catalogue should be interpreted as a partial geometric representation of facial phenotype rather than a complete translation of HPO descriptions. Individual measurements can be inspected through their landmark definitions and used in cohort-level statistical analyses, but FaceKit is not intended to determine the presence of an HPO term or provide a disease diagnosis.

The released generator reproduced the geometric phenotype of its training cohorts, both within a cohort and in what separates one cohort from the others, but the agreement depended on how many patients a cohort contributed. Within-cohort differences between synthetic and real measurements were smallest in the cohorts with 100 or more training patients and largest in the four cohorts with 13 to 19, and the correlation between synthetic and real cohort effects fell from 0.945 and 0.835 in the two largest tiers to 0.604 in the smallest. This decline cannot be separated from sampling noise here, because the smallest cohorts also contribute the fewest real and synthetic images to each comparison. Either way, the phenotype of a synthetic cohort is least reliable for the rarest syndromes, which are the ones for which augmentation would be most useful, and synthetic images of such a cohort should be checked against its real images before they are used in its place.

The privacy assessment supports distributing the pre-trained generator, but with a distinction that should be preserved when the synthetic images are used. On the identity axis, sampling from the audited generator was no more likely to produce a face close to a training patient than drawing another real photograph of the same syndrome was, and this held for all three recognition models at every operating point examined. It also did not depend on how many patients a cohort contributed to training: across 50 cohorts with 13 to 242 training patients, the proximity of synthetic images to the training partition showed no association with training-patient count, and no tier of cohorts exceeded the held-out reference. The expectation that a conditional generator reproduces individuals most readily in its smallest cohorts is therefore not supported at the cohort sizes examined here, although the four smallest cohorts give wide intervals and cannot exclude an effect. The nearest neighbor adversarial accuracy indicated a slight shift of the synthetic images toward the training partition as a whole, with an interval excluding zero for AdaFace only, and this shift was not accompanied by any excess in the flagging analysis. On the appearance axis the synthetic images were closer to the training photographs than held-out real images were at every operating point from 2\% upward, while the corresponding adversarial accuracy difference did not exclude zero. This proximity reflects the acquisition characteristics the generator learned alongside facial morphology rather than the identity of any individual, since the perceptual metric cannot separate two photographs of one patient from photographs of two different patients. Two limits apply to the audit. First, it was carried out on a generator fitted to the training partition alone, because the released generator has seen every available image and leaves no held-out reference; the released generator shares its architecture and training schedule but is fitted to 4,559 rather than 3,602 images, so its privacy is inferred from the audited generator rather than measured on it. Second, the assessment is an empirical audit against specific recognition models and a specific perceptual metric, not a formal privacy guarantee: it bounds what these particular measurements can detect, and a stronger attack or a better embedding could in principle detect more.

Several limitations should be considered. First, the reliability of every geometric measurement depends on the accuracy of the underlying facial landmark detector. General-purpose detectors are primarily trained on normative faces and may place landmarks according to learned expectations of typical facial structure when applied to rare-disease faces with substantially different morphology. Errors at this stage propagate directly into all downstream measurements and cannot be corrected through feature design alone. The same limitation applies to any phenotype defined by appearance rather than by shape, as the synophrys result above illustrates: a landmark mesh carries no photometric information and cannot represent hair distribution. Second, some feature failures arise from the choice of geometric formula or normalization scale, as in the full-cheeks result, indicating that normalization strategies may need to be selected separately for different measurements. Third, landmarks estimated from a single two-dimensional image remain sensitive to head pose, facial expression, image resolution, cropping, and acquisition conditions, despite the pose-quality filtering and geometric normalization implemented in FaceKit. The resolution analysis puts a number on one of these: the catalogue is stable to about 128 pixels of short side and acquires systematic offsets rather than merely added noise below it, so a measurement taken from a small image should not be compared on the same scale as one taken from a large one. The reference analyses put a number on another: the normative baseline depends on ancestry, and a reference conditioned on it recovers effects that the pooled reference misses, whereas conditioning on age or on image resolution changes nothing that can be distinguished from drift. These measurements should therefore not be interpreted as equivalent to calibrated three-dimensional facial measurements.

Future development will focus on supporting landmark-detection models adapted to rare-disease faces, refining feature-specific normalization strategies, and incorporating complementary appearance-based representations for phenotypes that cannot be described by geometry alone. Validation in larger cohorts and across demographic and imaging subgroups will also be needed before quantitative ranges can be interpreted clinically; the ancestry-invariance of the disease effect reported here rests on a single syndrome and on ancestry groups of 31 to 352 patients, and the smaller of those groups cannot exclude effects of the size that would matter clinically.

\section{Conclusion}

FaceKit converts a frontal facial photograph into 120 interpretable geometric
measurements and reports each of them on a normative scale, so that a facial
phenotype can be recorded as a set of quantities whose definitions are open to
inspection rather than as a free-text description or the output of a model that
cannot be examined. On a curated subset of the GestaltMatcher Database, the
measurement each HPO term is mapped to deviated from a healthy reference in the
direction the term predicts for 62 of the 84 assessable terms, and the cases
that did not are traceable to identifiable causes: a cohort-wide offset in the
measurement itself, a normalization scale altered by the phenotype it is meant
to measure, or a phenotype that resides in appearance rather than in shape. The
comparison of pose-correction strategies shows that the depth the correction
requires is better taken from a fixed canonical face than from the per-image
depth the landmark detector returns, the latter lowering intra-patient
reliability below the level obtained with no correction at all. The generator
distributed with the toolkit reproduces the geometric phenotype of the cohorts
it was trained on, both within a cohort and in what separates one cohort from
another, with a modest narrowing of the variation inside a cohort and weaker
agreement in the cohorts with the fewest training patients; and a generator
trained in the same way on the training partition produces outputs that carry no
identity signal in excess of the rate that real held-out photographs of the same
cohorts produce, irrespective of how many patients a cohort contributed.

Taken together, these results position FaceKit as an instrument for measurement
rather than for diagnosis. A FaceKit measurement is a reproducible quantity with
an explicit landmark definition and a stated reference, suitable for
cohort-level comparison, for genotype--phenotype analysis, and as a structured
input to downstream models; it is not a determination that an HPO term is
present, and the quantitative ranges reported here are not yet calibrated for
clinical interpretation. Reaching that point will require reference
distributions validated across a wider range of ancestry, age, and acquisition
conditions than the present controls span, alongside the improvements to
landmark detection and to normalization set out above.

\section{Competing Interests}
The authors declare that they have no competing interests.

\section{Acknowledgment}
We acknowledge the publicly available data resources used in this work, including the GestaltMatcher Database (GMDB) and RDFace, and thank the respective dataset creators, curators, and maintainers for making these resources available. We also acknowledge the use of internal data resources from Children's Hospital of Philadelphia (CHOP) and thank the CHOP Arcus team for their support in facilitating data access, management, and computational infrastructure. This work was supported by NIH grant OD037960 and the CHOP Research Institute.

\section{Author's Contribution}
H.C. and K.W. conceived the study. H.C. and F.P. implemented the method and performed experiments. H.C. and Z.W. analyzed results and wrote the manuscript. J.B. tested the software. T.C.H., P.K. provided dataset. M.U.A., I.M.C., C.L., W.K.C., C.W., G.G. advised on the study. K.W. guided the execution of the study. 

\bibliographystyle{unsrt}
\bibliography{references}

\end{document}